\documentclass[a4paper]{elsarticle} % arXiv: set a4paper

\journal{arXiv.org}
\usepackage[footnotesize]{caption}  % arXiv: set caption font size
\usepackage[colorlinks=true]{hyperref}
\usepackage[T1]{fontenc}
\usepackage{bookmark}
\usepackage{xcolor}

\usepackage{amsfonts}
\usepackage{amsmath}
\usepackage{amssymb}

\usepackage{graphicx}
\usepackage{algorithm}
\usepackage{algorithmicx}
\usepackage{algpseudocode}
\usepackage{subcaption}
\usepackage{listings}

\algnewcommand \Input[3]  {\item\hspace*{0ex}\textbf{input}  #1 : #2 \Comment{#3}}
\algnewcommand \Output[3] {\item\hspace*{0ex}\textbf{output} #1 : #2 \Comment{#3}}

\newcommand{\MPI}[1]{\textsf{\detokenize{MPI_#1}}}

\newcommand{\bs}[1]{\boldsymbol{#1}}

\begin{document}

\begin{frontmatter}

  \title{Fast parallel multidimensional FFT using advanced MPI}

  \author[KAUST]{Lisandro Dalcin}\ead{dalcinl@gmail.com}

  \author[UiO]{Mikael Mortensen\corref{corauthor}}\ead{mikaem@math.uio.no}

  \author[KAUST]{David E. Keyes} \ead{david.keyes@kaust.edu.sa}

  \address[KAUST]{%
    Extreme Computing Research Center\\
    King Abdullah University of Science and Technology\\
    Thuwal, Saudi Arabia\\
    \url{https://ecrc.kaust.edu.sa}}

  \address[UiO]{%
    Department of Mathematics\\
    University of Oslo,
    Oslo, Norway\\
    \url{http://www.mn.uio.no/math/english/}}

  \cortext[corauthor]{Corresponding author}

  \begin{abstract}
    We present a new method for performing global redistributions of
    multidimensional arrays essential to parallel fast Fourier (or
    similar) transforms. Traditional methods use standard all-to-all
    collective communication of contiguous memory buffers, thus
    necessary requiring local data realignment steps intermixed
    in-between redistribution and transform steps. Instead, our method
    takes advantage of subarray datatypes and generalized all-to-all
    scatter/gather from the MPI-2 standard to communicate
    discontiguous memory buffers, effectively eliminating the need for
    local data realignments. Despite generalized all-to-all
    communication of discontiguous data being generally slower, our
    proposal economizes in local work. For a range of strong and weak
    scaling tests, we found the overall performance of our method to
    be on par and often better than well-established libraries like
    MPI-FFTW, P3DFFT, and 2DECOMP\&FFT. We provide compact routines
    implemented at the highest possible level using the MPI bindings
    for the C programming language. These routines apply to any global
    redistribution, over any two directions of a multidimensional
    array, decomposed on arbitrary Cartesian processor grids (1D
    slabs, 2D pencils, or even higher-dimensional decompositions). The
    high level implementation makes the code easy to read, maintain,
    and eventually extend. Our approach enables for future speedups
    from optimizations in the internal datatype handling engines
    within MPI implementations.
  \end{abstract}


\end{frontmatter}

\section{Introduction}
\label{sec:introduction}

The Fast Fourier Transform (FFT) remains one of the most significant
algorithms across various disciplines in science and
society. Applications range from image analysis and signal processing
to the solution of partial differential equations through spectral
methods. Spectral methods are frequently the method of choice for
physicists that aim for the most accurate numerical methods to get
representations of physical models as realistic as possible. In
particular, FFT-based spectral methods are at the core of all major
Direct Numerical Simulation (DNS) codes used in fundamental studies of
turbulence and transitional flows. These simulations are pushing the
limits of high-performance supercomputers, with computational domains
approaching trillions of unknowns. Within such applications, it is
crucial to ensure the best possible algorithms for both serial and
parallel FFT, which often is the bottleneck of the codes.

It is well known that an FFT on multidimensional data can be performed
as a sequence of one-dimensional transforms along each dimension.  For
example, a multidimensional array of shape $N_x \times N_y \times N_z$
can be Fourier-transformed by first performing $N_x \times N_y$ serial
transforms of length $N_z$ along the last axis, followed by
$N_x \times N_z$ transforms of length $N_y$ along the middle axis and
then finally $N_y \times N_z$ transforms of length $N_x$ along the
first axis. However, when the computational domains become too large
to fit in the memory locally available within a single compute unit,
the domain have to be distributed amongst several, often thousands. In
this case, only a small part of the multidimensional array is
available on each processor.

It is the job of decomposition algorithms and global redistribution
operations to assist in the computation of the multidimensional
FFT, by ensuring that array data needed for a serial 1D transform
along a given axis is locally available when needed. In the
literature, such approaches are often referred as \emph{transpose}
algorithms~\cite{foster97}.  An alternative parallel FFT method, which
is more intrinsically connected with the FFT algorithm, is the binary
exchange (or \emph{distributed}) method. In this work we will only
consider the transpose algorithm, which in general is found to be
superior for large problems, and refer to Gupta and
Kumar~\cite{gupta93} and Foster and Worley~\cite{foster97} for a
review of both methods.

At first, multidimensional parallel FFTs based on global
redistributions were conducted using slab decompositions, where only
one axis of a multidimensional array is distributed.  Despite slab
decompositions being very efficient, they are unfortunately limited to
a rather small number of processors, since that number cannot be
larger than $N$, assuming $N=N_x=N_y=N_z$. The next level of
parallelism was reached with 2D pencil decompositions \cite{Ding95},
where two axes of a multidimensional array were distributed, using
one-dimensional subgroups of processors corresponding to rows and
columns of a logically two-dimensional processor grid. Pencil
decompositions are usually found to be less efficient than slab
decompositions, but the number of processors can be as large as $N^2$.
For such reason, pencil decompositions became the only sensible choice
for large-scale simulations using hundreds of thousands of processors.

Several open-source implementations of parallel FFT based on global
redistributions are available. For pencil decompositions the
P3DFFT~\cite{pekurovsky2012} and 2DECOMP\&FFT~\cite{Li2010} are
probably the two most commonly used libraries, both being implemented
in Fortran 90 and using similar algorithms based on collective
all-to-all communication of contiguous arrays followed (or preceded)
by a local transpose or remapping operation.  Both libraries
primarily target three-dimensional arrays and complex-to-complex or
real-to-complex/complex-to-real transforms.  The PFFT package of
Pippig~\cite{Pippig13} is more general and can be used for even
higher-dimensional arrays, using processors grids with more than two
dimensions. PFFT is built on top of FFTW \cite{fftw}, which comes with
its own slab implementation. However, instead of using FFTW's built-in
slab implementation, PFFT makes use of FFTW's global transpose
routines to implement the 2D (or even higher dimensional) pencil
method. Other known parallel FFT libraries are OpenFFT~\cite{DUY2014},
which also admits higher ($>3$) dimensional transforms,
AccFFT~\cite{accfft}, which utilizes both CPUs and GPUs, the parallel
FFT subroutine library of Plimpton~\cite{plimptonFFT},
FFTW++~\cite{fftwpp}, which implements both binary exchange and
transpose algorithms, and mpiFFT4py~\cite{mpifft4py}, which provides a
high-level Python interface based on MPI for
Python~\cite{mpi4py,dalcin08}.

In this paper we suggest a completely generic, black-box, global
redistribution method, based on the generalised all-to-all
(\MPI{ALLTOALLW}) scatter/gather and subarray datatype facilities
available in the Message Passing Interface (MPI)
standard~\cite{mpistd31}.  To the best of our knowledge this approach
has not been explored much in the literature. Derived datatypes were
used previously for a slab decomposition by Hoefler and Gottlieb
\cite{hoefler10}, where the local transposes used by regular parallel
MPI implementations were described as part of the derived datatype,
and speedup over traditional algorithms was demonstrated for some, but
not all showcases. The Warp~\cite{warp} particle-in-cell code is using
\MPI{ALLTOALLW} with derived datatypes for global redistributions, but
only for 3D arrays and power-of-two number of processors.

It is in the spirit of extreme-scale architecture design to
restructure algorithms to allow taking the rearrangement of data off
the critical path of the CPU and into the memory subsystem or the
network, provided that the supporting hardware and software of those
layers can accommodate. The global redistribution method described in
this paper embraces such a paradigm shift, as it eliminates the need
for any local remappings. Furthermore, it is
applicable to arrays of arbitrary dimensions, decomposed on Cartesian
processor grids that are also of arbitrary dimensionality. As in
previous parallel (transpose) FFT implementations we assume that there
is a serial FFT code already available, and discuss only the parallel
decomposition and collective communication required to utilise such a
serial code most efficiently in parallel. To this end we introduce
some necessary theory and notation on discrete Fourier transforms in
Sec~\ref{sec:seqfft}.  In Sec~\ref{sec:mpi}, some existing FFT
transpose methods are discussed before introducing our new global
redistribution method. In Sec \ref{sec:results}, we compare scaling of
the new method with other well-known parallel FFTs libraries on a Cray
XC40 supercomputer. Finally, conclusions are drawn in
Sec~\ref{sec:conclusions}.

\section{Sequential FFTs of multidimensional arrays}
\label{sec:seqfft}

The discrete Fourier transform (DFT) takes a sequence of complex
numbers $u_0, u_1, \ldots, u_{N-1}$ and transforms them into another
sequence of complex numbers
$\hat{u}_0, \hat{u}_1, \ldots, \hat{u}_{N-1}$. Forward and backward
transforms can be defined, respectively, as
\begin{align}
\hat{u}_k &= \frac{1}{N} \sum_{j=0}^{N-1} {u}_j e^{-ikx_j} \quad k=0, 1, \ldots, N-1, \\
u_j &= \sum_{k=0}^{N-1}\hat{u}_{k} e^{ikx_j} \quad j=0, 1, \ldots, N-1,
\end{align}
where $i$ is the imaginary unit, $x_j = 2\pi j/N$, and simplifications
are possible if either sequence is real. An alternative and more
compact notation is
\begin{align}
\bs{\hat{u}} &= \mathcal{F}(\bs{u}), \\
\bs{u} &= \mathcal{F}^{-1}(\bs{\hat{u}}),
\end{align}
where $\bs{u} = \{u_j\}_{j=0}^{N-1}$,
$\bs{\hat{u}} = \{\hat{u}_k\}_{k=0}^{N-1}$ and
$ \bs{u} = \mathcal{F}^{-1}(\mathcal{F}(\bs{u}))$.

The forward and backward DFTs are usually computed using a fast
Fourier transform (FFT) algorithm. In this work we will assume that
there exist high-performance, serial (single-process, maybe
multi-threaded) FFT routines to compute these one-dimensional forward
and backward DFTs. These routines are widely available from, e.g.,
FFTW \cite{fftw}, FFTPACK \cite{fftpack}, IBM~ESSL \cite{ibm-essl}, or
Intel~MKL \cite{intel-mkl}.

In many applications numerical data are arranged in multidimensional
arrays. We denote a $d$-dimensional array as
$u_{j_0, j_1, \ldots, j_{d-1}} $, where there are $d$ index sets
$j_m=0,1,\ldots,N_m-1$, $m \in 0,1,\ldots, d-1$, with $N_m=|j_m|$
being the length of $j_m$. A forward $d$-dimensional DFT on the
$d$-dimensional array ${u}_{j_0, j_1, \ldots, j_{d-1}}$ will then be
computed as
\begin{equation}
\hat{u}_{k_0, k_1, \ldots, k_{d-1}} = \sum_{j_0=0}^{N_0-1}
\left( \frac{\omega_{0}^{k_0 j_0}}{N_0} \sum_{j_1=0}^{N_1-1}
\left( \frac{\omega_{1}^{k_1 j_1}}{N_1}  \cdots \sum_{j_{d-1}=0}^{N_{d-1}-1}
\frac{\omega_{{d-1}}^{k_{d-1} j_{d-1}}}{N_{d-1}}  {u}_{j_0, j_1, \ldots, j_{d-1}}
 \right) \right),
\label{eq:multifft}
\end{equation}
where $\omega_{j} = e^{-2 \pi i /N_j}$. Note that for a transformed
axis $m$ we use the index set $k_m = 0, 1, \ldots N_m-1$ instead of
$j_m$, and (\ref{eq:multifft}) is executed over all these output
indices\footnote{If the transform involves a real sequence, then
simplifications are possible due to Hermitian symmetry, and we can
use the smaller index set $k_{d-1}=0,1,\ldots, N_{d-1}/2$. Also note
that it is perhaps more common to use transformed index sets centered
around zero, like $k_m = -N_m/2, -N_m/2+1, \ldots, N_m/2-1$.}.
As such, we see in (\ref{eq:multifft}) that the array
$\hat{u}_{k_0, k_1, \ldots, k_{d-1}}$ has been transformed along all
of its axes. Note that a hat notation, $\hat{u}$, is used exclusively
for a fully transformed array, i.e., the output of a complete forward
FFT over all axes.

We can simplify Eq.~(\ref{eq:multifft}) using the notation
\begin{equation}
\hat{u}_{k_0, k_1, \ldots, k_{d-1}} =
\mathcal{F}_0 \left(
\mathcal{F}_1 \left( \cdots
\mathcal{F}_{d-1}({u}_{j_0, j_1, \ldots, j_{d-1}})
\right)
\right),
\label{eq:fftn_forward}
\end{equation}
where $\mathcal{F}_i(\cdot)$ represents a partial transform, i.e., a
one-dimensional DFT along axis $i$, for all other index sets unchanged
\begin{equation}
\tilde{u}_{j_0, \ldots, k_i, \ldots, j_{d-1}} =
\mathcal{F}_i(u_{j_0, \ldots, j_i, \ldots, j_{d-1}}).
\end{equation}
Note that this represents exactly $1/N_i\prod_{m=0}^{d-1} N_m $
one-dimensional transforms of length $N_i$. Also note that here, and
for the rest of this paper, tilde notation, $\tilde{u}$, is used to
represent an array that is only partially transformed, i.e.,
transformed along some, but not all, of its axes.

From Eq.~(\ref{eq:fftn_forward}) it is evident that the DFTs are
computed in sequence, one axis of the multidimensional array at a
time. A backward $d$-dimensional DFT is executed in the opposite order
\begin{equation}
{u}_{j_0, j_1, \ldots, j_{d-1}} =
\mathcal{F}^{-1}_{d-1} \left( \cdots
\mathcal{F}^{-1}_1 \left(
\mathcal{F}^{-1}_{0} (\hat{u}_{k_0, k_1, \ldots, k_{d-1}})
\right)
\right).
\label{eq:fftn_backward}
\end{equation}

Moving from one to several dimensions, the data arrays quickly grow
in size, and it becomes necessary to distribute the arrays across
several processors within distributed-memory computing
architectures. Since the multidimensional FFTs are computed in
sequence, one axis at a time, we need only ensure that the
whole length of the array along that one axis is available on each
single processor when it is up for transformation. Making this happen is
the job of global array redistribution procedures, using parallel
decompositions and communication as discussed briefly in the
introduction, and in sections to come.

\section{Parallel FFTs of multidimensional arrays}
\label{sec:mpi}

Consider a $d$-dimensional array $u_{j_0, j_1, \ldots, j_{d-1}}$ and
pick any one of the index sets $j_m=0,1,\ldots,N_m-1$,
$m \in 0,1,\ldots,d-1$. This index set can be partitioned (and
corresponding array entries mapped) into an ordered group of processes
$P$ of size $|P|$ with process identifiers
$p=0,1,\ldots,|P|-1$. Regardless of how it is partitioned, we denote
an index set $j_m$ distributed into a process group $P$ as $j_m/P$. As
such, a $d$-dimensional array that is distributed in its first axis by
processor group $P$ will be denoted as
$u_{j_0/P, j_1, \ldots, j_{d-1}}$. Note that here and throughout this
paper we assume that arrays are in C-style
row-major order. For Fortran-style column-major order, it would be
natural to distribute the last index set $j_{d-1}$ rather than $j_0$.

\subsection{Balanced block-contiguous decompositions}
\label{sec:decomp}

There are many different ways of distributing the index set $j_m$ on a
process group $P$. From the many choices available, we restrict our
discussion to block-contiguous decompositions. Such decompositions are
fully defined from the global index set length $N_m=|j_m|$, the number
of processes $M=|P|$, and local (that is, within each processor) index
set lengths. We denote these local index set lengths of $j_m$ as
$N_m/P$, they correspond to a sequence $\{(N_m/P)_p\}_{p=0}^{M-1}$.
For simplicity, and only in this section, we will use the notation
$n_p=(N_m/P)_p$ to refer to the local length corresponding to the
$p$-th process.  Within the obvious restriction $\sum_p n_p = N_m$,
the values $n_p$ are otherwise arbitrary. In practice, it is useful to
compute and store the start index $s_p$ corresponding to each process
with the recursion $s_0 = 0$, $s_k = s_{k-1} + n_{k-1}$,
$k=1,2,\ldots,M-1$.

A balanced block-contiguous decomposition of a sequence of $N_m$
elements in $M$ parts $p=0,1,\ldots,M-1$ is given by the simple
formula %
\footnote{To the best of our knowledge this formula was introduced
by Barry Smith in the 90's as part of the fundational development
of PETSc~\cite{petsc-user-ref}. Since then, this
formula has been the default decomposition strategy for distributed
vectors and matrices.}
\begin{align}
n_p =
\begin{cases}
q+1 & \text{if } r > p \\
q & \text{otherwise}
\end{cases}, \text{ with }
q = \left\lfloor\frac{N_m}{M}\right\rfloor \text{ and }
r = N_m \bmod M.
\label{eq:decompose}
\end{align}

Alg.~\ref{alg:Decompose} shows pseudocode using
Eq.~(\ref{eq:decompose}) to compute the local lengths $n_p$ along with
an explicit, non-recursive expression for the start indices $s_p$. %
Executing the call
$n_p,s_p\leftarrow\text{\textsc{Decompose}}(|j_m|,|P|,p)$ implicitly
defines an subset $\{s_p,\ldots,s_p+n_p-1\}$ of $j_m$ corresponding to
the $p$-th process in group $P$. %
Although admittedly trivial, for the sake of completeness we present
in Listing~\ref{lst:Decompose} a concrete and concise implementation
in the C programming language. For the rest of this paper we trade
generality for simplicity and restrict our discussion to balanced
block-contiguous decompositions as defined in
Eq.~(\ref{eq:decompose}). Other pseudocodes and listings to be presented
later are greatly simplified, as the various $n_p$, $s_p$ values can
be computed with Alg.~\ref{alg:Decompose} on the fly and as needed
rather than having to store and pass them around function calls.

\begin{algorithm}[h!]
\caption{Balanced block-contiguous decomposition}
\label{alg:Decompose}
\begin{algorithmic}[1]
\Function{Decompose}{$N, M, p$}
\Input{$N$}{integer}{total number of elements, $N \geq 0$}
\Input{$M$}{integer}{number of parts, $M > 0$}
\Input{$p$}{integer}{part index, $0 \leq p < M$}
\Output{$n$}{integer}{number of elements in $p$-th part}
\Output{$s$}{integer}{start index of $p$-th part}
\State $q \gets \lfloor N/M \rfloor $
\State $r \gets N\bmod M$
\If{$r > p$}
\State $n \gets q + 1$
\State $s \gets n \cdot p$
\Else
\State $n \gets q$
\State $s \gets n \cdot p + r$
\EndIf
\State \textbf{return} $n, s$
\EndFunction
\end{algorithmic}
\end{algorithm}

\begin{lstlisting}[float=h!,frame=single,
caption={Balanced block-contiguous decomposition},
label=lst:Decompose,
]
#define min(x, y) (((x) < (y)) ? (x) : (y))
void decompose(int N, int M, int p, int *n, int *s)
{
  int q = N / M;
  int r = N % M;
  *n = q + (r > p);
  *s = q * p + min(r, p);
}
\end{lstlisting}

\subsection{Global redistributions}
\label{sec:global_redist}

We can perform a serial FFT on any index set of a multidimensional
array that is not distributed. For example, for the array
${u}_{j_0/P, j_1, \ldots, j_{d-1}}$, we can perform a partial
transform over all but the first axis as
\begin{equation}
\tilde{u}_{j_0/P, k_1, \ldots, k_{d-1}} =
\mathcal{F}_1 \left(
\mathcal{F}_2 \left( \cdots
\mathcal{F}_{d-1}({u}_{j_0/P, j_1, \ldots, j_{d-1}})
\right)
\right).
\label{eq:rfftn_slab}
\end{equation}
However, we cannot perform the transform over the first axis, because
only a part of the global array is available locally on each
process. It is the job of global redistribution (or \emph{transpose})
operations to ensure that data within a distributed array is realigned
such that a distributed axis becomes locally available in full for all
processes in the group.  We denote a global redistribution operation
from alignment in axis $v$ to alignment in axis $w$ performed within a
process group $P$ as
\begin{equation}
u_{\ldots, j_w, \ldots, j_v/P, \ldots}
\xleftarrow[P]{v \rightarrow w}
u_{\ldots, j_w/P, \ldots, j_v, \ldots}.
\label{eq:general_gt}
\end{equation}
Note that axes other than $v$ and $w$ are not involved in the
redistribution operation and thus they are not altered by the
exchange.

The global redistribution operation brings us to the main novelty of
this paper. All known parallel FFT libraries perform global
redistributions in two steps (not necessarily in this order):%
\vskip 0.1in%
\begin{itemize}
\item[1)] Contiguous data communication using collective all-to-all
  operations.
\item[2)] Local data rearrangements or transpose operations, also
  referred as remappings.
\end{itemize}
\vskip 0.1in%
Both steps are known to be computationally expensive. The first step
involves communicating large amounts of data among processes within
the group in an all-to-all fashion. The second step involves
non-contiguous memory accesses and copies, which is heavily affected
by cache capacity and memory bandwidth of current computing
architectures. %

In this work, we suggest a global redistribution method that
eliminates the need for any local remappings or transposes. To explain
how our method works, and how it differs from other methods, we
discuss first in Sec.~\ref{sec:old_redist} a
traditional implementation of parallel FFTs with slab decomposition
using local remappings and all-to-all communication of contiguous
memory buffers. Afterwards, in Sec~\ref{sec:new_redist}, we show how
the slab decomposition can be implemented without local remappings
using subarray datatypes and generalized all-to-all
communication. We will then, in Sec~\ref{sec:cartesian}, briefly
describe multidimensional Cartesian process topologies, that are
to be utilized in Secs.~\ref{sec:pencil} and \ref{sec:4d},
where the approach will be shown to extend trivially to the 2D
pencil method and even higher-dimensional processor grids.

\subsection{Slab decomposition}
\label{sec:slab}

With the notation introduced previously, a parallel FFT on a
multidimensional array $u_{j_0, j_1, \ldots, j_{d-1}}$, that is
initially distributed in a processor group $P$ in the first index set
$j_0$, can be performed in three steps:
\begin{align}
\tilde{u}_{j_0/P, k_1, \ldots, k_{d-1}} &= \mathcal{F}_1\left( \mathcal{F}_2\left(  \ldots \mathcal{F}_{d-1}(u_{j_0/P, j_1, \ldots, j_{d-1}})  \right)\right),\label{eq:slab0} \\
\tilde{u}_{j_0, k_1/P, \ldots, k_{d-1}} &\xleftarrow[P]{1\rightarrow 0} \tilde{u}_{j_0/P, k_1, \ldots, k_{d-1}}, \label{eq:gt_slab} \\
\hat{u}_{k_0, k_1/P, \ldots, k_{d-1}} &= \mathcal{F}_0(\tilde{u}_{j_0, k_1/P, \ldots, k_{d-1}}).\label{eq:slab1}
\end{align}
These steps correspond to a traditional parallel FFT with slab
decomposition, as illustrated in Fig.~\ref{fig:slab} for a
three-dimensional array distributed in a group of four processes.  We
are interested in the global redistribution in the second step, that
is usually accomplished with a local remapping
followed by a collective all-to-all communication.

\subsubsection{Traditional global redistribution method}
\label{sec:old_redist}

\begin{figure}[t!]
  \centering
  \begin{subfigure}[t]{0.45\textwidth}
    \centering
    \includegraphics[width=\textwidth]{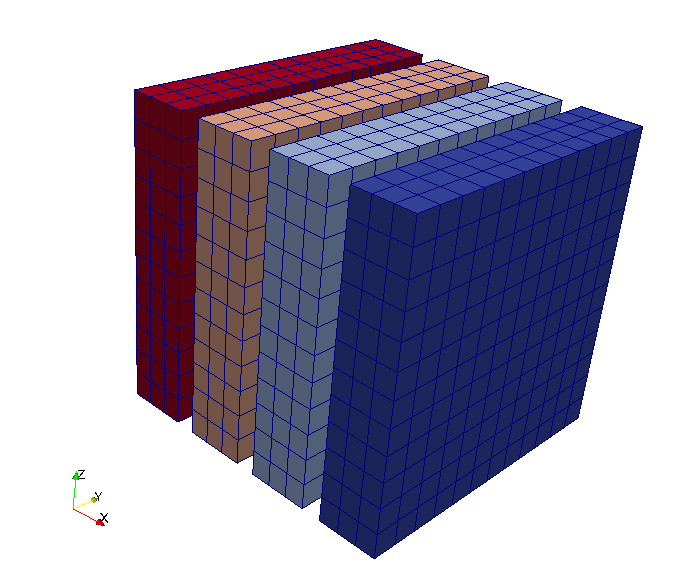}
    \caption{Global array distributed in the $x$-direction.}
    \label{fig:slab:a}
  \end{subfigure}
  \begin{subfigure}[t]{0.45\textwidth}
    \centering
    \includegraphics[width=\textwidth]{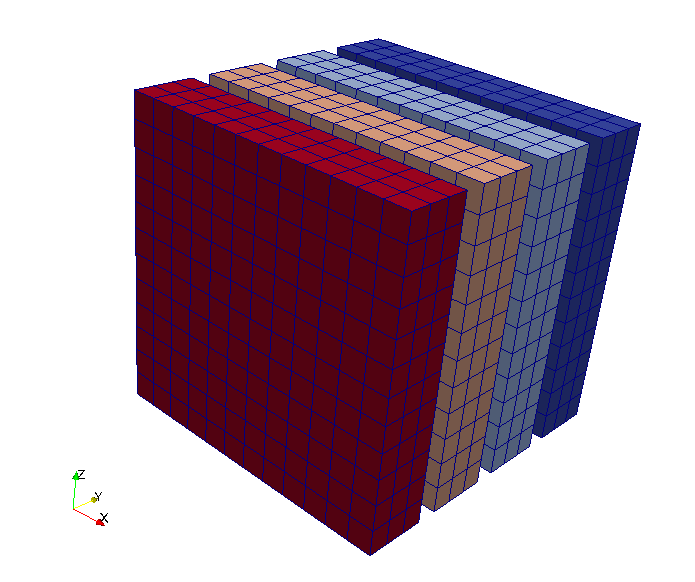}
    \caption{Global array distributed in the $y$-direction.}
    \label{fig:slab:b}
  \end{subfigure}
  \caption{Slab decomposition. Colours represent CPU rank. Red,
    orange, light blue and dark blue represent CPUs 0, 1, 2, and 3
    respectively.}
  \label{fig:slab}
\end{figure}

\begin{figure}[h!]
  \centering
    \begin{subfigure}[c]{0.33\textwidth}
        \centering
        \includegraphics[scale=0.27]{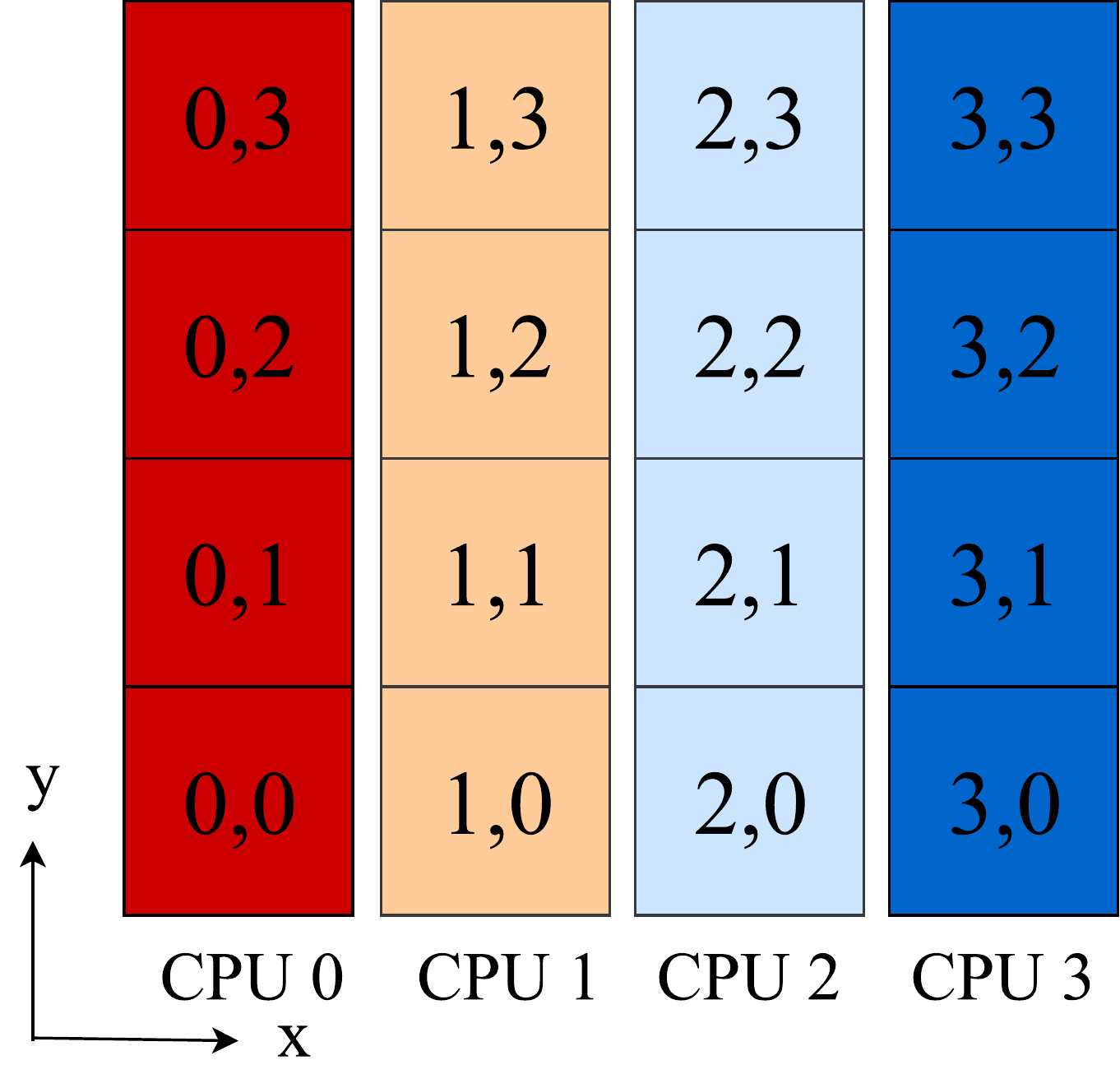}
        \caption{Original array.}
        \label{fig:slab2D:a}
    \end{subfigure}%
    \begin{subfigure}[c]{0.33\textwidth}
        \centering
        \includegraphics[scale=0.27]{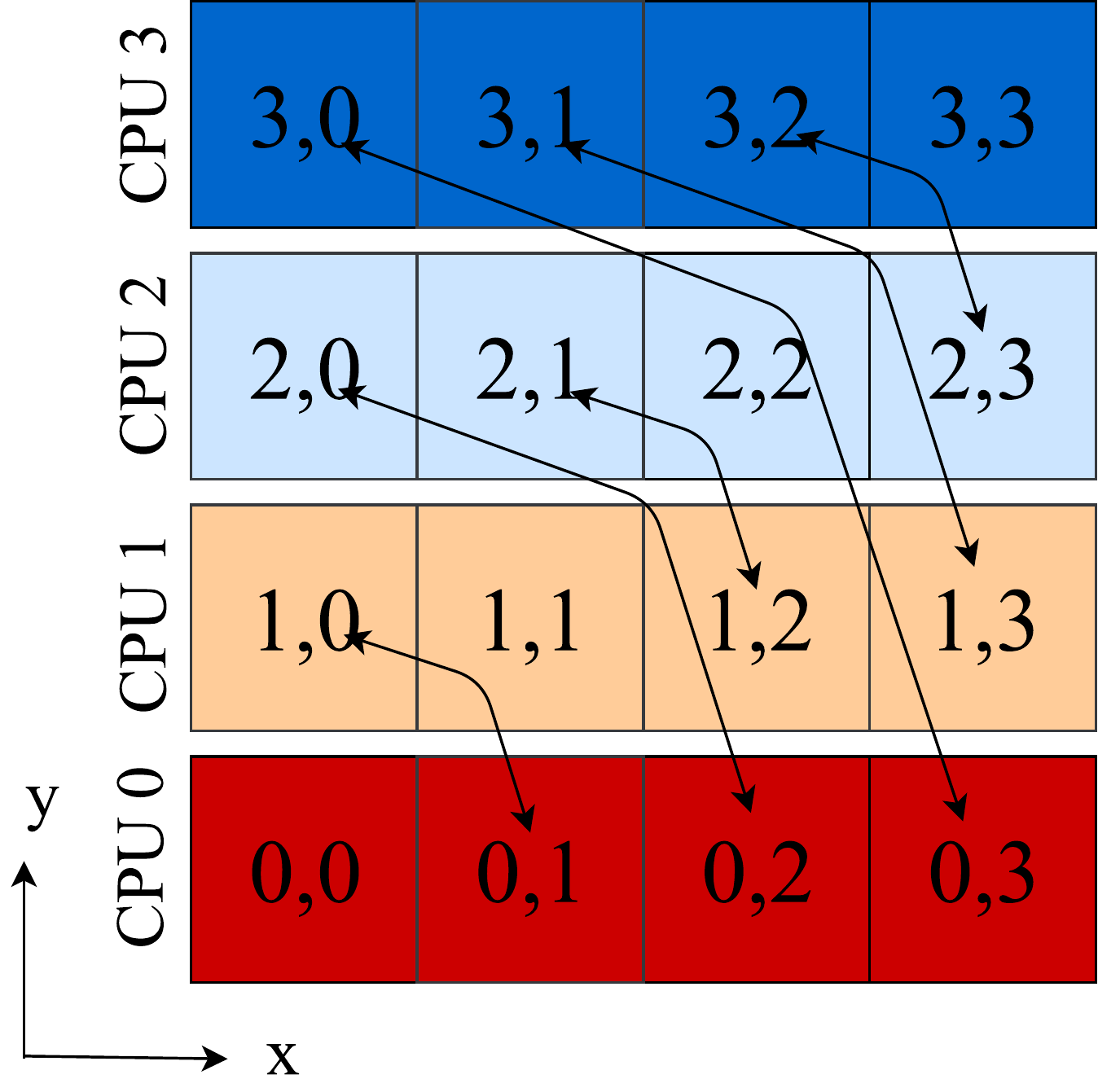}
        \caption{After local transpose.}
        \label{fig:slab2D:b}
    \end{subfigure}
    \begin{subfigure}[c]{0.33\textwidth}
        \centering
        \includegraphics[scale=0.27]{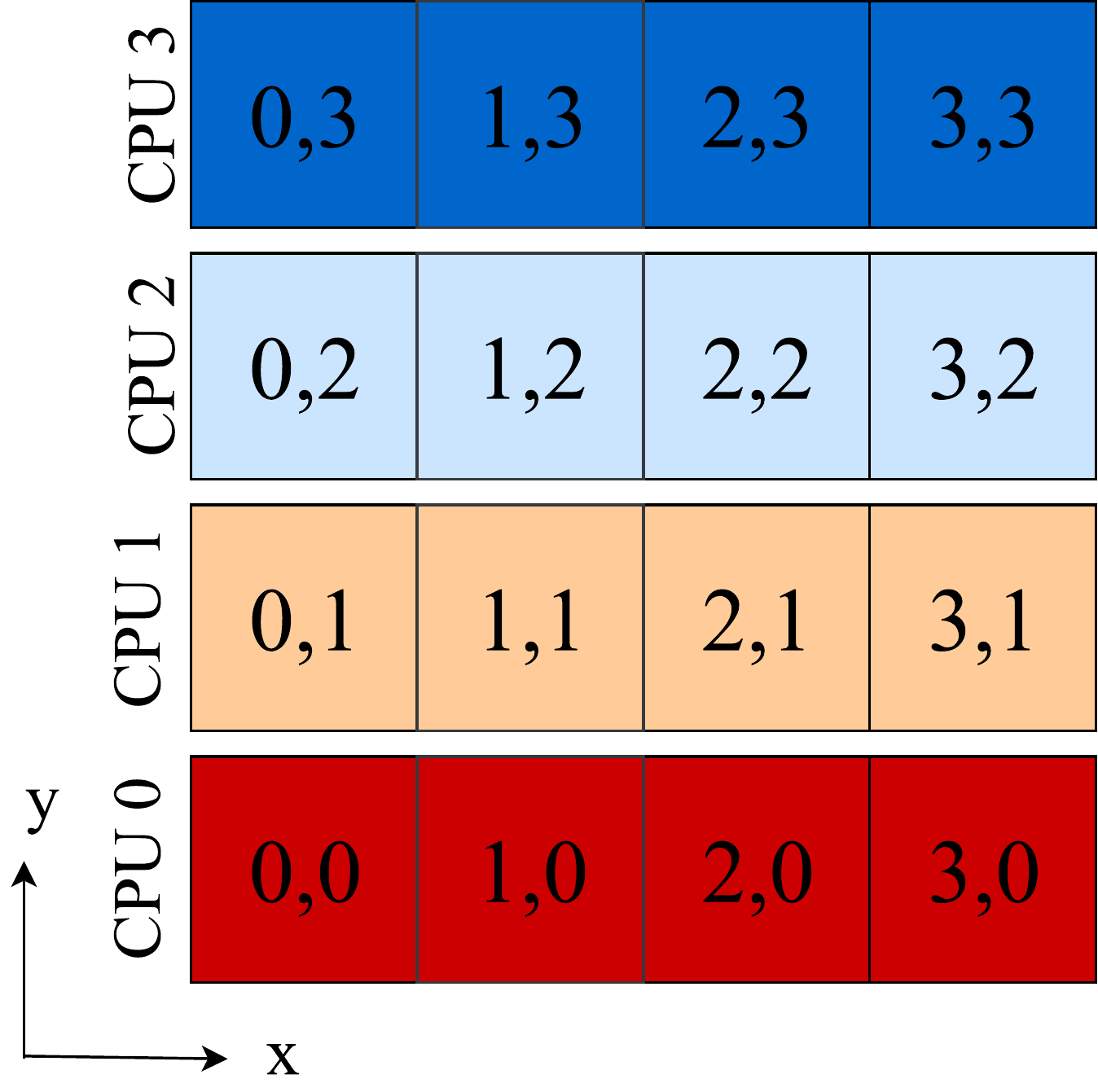}
        \caption{After Alltoall.}
        \label{fig:slab2D:c}
    \end{subfigure}
  \caption{Illustration of the slab decomposition. %
    In (a) we see the original global array from Fig~\ref{fig:slab:a}
    projected on the $xy$-plane. Each slab is divided into 4 smaller
    chunks that are to be communicated with the other processors and
    each chunk is identified with a pair of integer labels. %
    In (b) we see the layout for slabs and chunks after performing a
    local transpose to align data in $x$-direction.  Bidirectional
    arrows represent a subsequent all-to-all exchange among
    processors. %
    In (c) we see the layout of the final global array after
    redistribution. Note that the relative position (within the global
    array) of each chunk in (a) is preserved in (c), however most
    chunks have migrated to a different processor.}
  \label{fig:slab2D}
\end{figure}

For a three-dimensional array $\tilde{u}_{j_0/P, k_1, k_2}$ (see \ref{eq:gt_slab}),
we can illustrate a global redistribution based on \MPI{ALLTOALL}
on a projected $xy$-plane, since the third index set $k_2$ is not
affected by the exchange.  Fig.~\ref{fig:slab2D} is an illustration of
Fig.~\ref{fig:slab} as seen along the $z$-axis, and with the local
arrays divided into chunks (or subarrays), four chunks for each slab
since there are four processes. Each of the chunks within a process has
to be communicated with the other processes. The chunks are labelled
with processor number first, and then chunk number. Now, to be able to
perform an all-to-all communication the local arrays as seen in
Fig.~\ref{fig:slab2D:a} need to be packed in a contiguous array
such that the chunk going out to rank 0 comes first in memory, then
the chunk that goes out to rank 1, and so on.
In other words, the local arrays must be remapped to an $x$-alignment
with shape $(N_0, N_1/P, N_2)$, as seen in Fig.~\ref{fig:slab2D:b}
\footnote{Note that a array of shape $(N_0, N_1/P, N_2)$ in
row-major order is laid out in memory exactly as an array of
shape $(P, N_0/P, N_1/P, N_2)$, it merely has one less index set and
stride.}. This operation is usually referred as a transpose, or
permutation, and it is local to each processor. Assuming here
for simplicity that $N_0$ and $N_1$ are divisible by $|P|$,
the local transpose operation on the $(N_0/P, N_1,N_2)$-shaped
array can be performed as follows (see Fig 1 of \cite{mortensen16})
\begin{align}
  (N_0/P, P, N_1/P, N_2) &\xleftarrow{\text{Reshape}}                       (N_0/P, N_1, N_2) \\
  (P, N_0/P, N_1/P, N_2) &\xleftarrow{\text{Swap axes } 0 \leftrightarrow 1} (N_0/P, P, N_1/P,N_2) \label{eq:transpose} \\
  (N_0, N_1/P,N_2)       &\xleftarrow{\text{Reshape}}                        (P, N_0/P, N_1/P,N_2)
\end{align}
Note that the first and last operations merely represent changes of
strides and index sets, and the cost is next to nothing. The transpose
operation (\ref{eq:transpose}), that swaps the first two axes, is the
costly part.  After transposing the local arrays to the shapes seen in
Fig.~\ref{fig:slab2D:b}, the datachunks that are to be communicated
are contiguous in memory and we may now simply call a collective
all-to-all, where the communication pattern is illustrated with
bidirectional arrows in Fig.~\ref{fig:slab2D:b}. The resulting arrays
are as shown in Fig.~\ref{fig:slab2D:c}.

Note that there are several different ways of performing a global
redistribution, and if the global array sizes are not divisible by
$|P|$, then the local transpose operation is more complex, and
\MPI{ALLTOALLV} must be used in place of \MPI{ALLTOALL}. FFTW provides
three global redistribution (termed global transpose by FFTW)
algorithms, where it is possible to choose a different stride on the
output and input arrays, and to combine this with serial FFTs on
non-contiguous data. Considering the 3D data in this section, the
action of FFTWs global redistribution (that includes all-to-all or
similar communication) is then either one of
\begin{align}
(N_0, N_1/P, N_2) &\xleftarrow{\text{Regular}}        (N_0/P, N_1, N_2), \label{eq:fftw_regular_out} \\
(N_1/P, N_0,N_2)  &\xleftarrow{\text{Transposed out}} (N_0/P, N_1,N_2).  \label{eq:fftw_transposed_out}
\end{align}
For FFTW the transposed out option is the fastest, since the regular
algorithm is using the transposed out algorithm followed by a global
redistribution. FFTW provides an interface that allows for
simultaneous planning of local array transpositions and serial FFTs in
one single step. PFFT takes advantage of these routines provided in
FFTW and performs planning for both the global redistribution and the
serial FFTs stages. The choice of making the axes of output arrays
transposed in reference to the input is made for efficiency, but
naturally it adds a level of complexity, and it is left to the user to
make sure that array operations on output arrays take this ordering
into consideration. This added complexity is also present in PFFT and
P3DFFT, these libraries have options to output arrays with either
regular or transposed alignment. The new global redistribution method,
to be described in the next subsection, does not transpose the axes of
input or output arrays.

\subsubsection{A new global redistribution method}
\label{sec:new_redist}

As discussed in Sec.~\ref{sec:old_redist}, global redistributions for
slab decompositions require two steps: i) a local remapping or
transpose to rearrange array data in contiguous memory buffers and ii)
collective all-to-all communication with these contiguous memory
buffers. In the following we describe how the same outcome can be
achieved in a single collective communication step. The approach is
straightforward if one relies on two slightly advanced
features introduced in the MPI-2.0 version of the standard more than
20 years ago. These features allow for collective all-to-all
communication of discontiguous memory buffers described through derived
datatypes, effectively eliminating the need for any local remapping
steps to ensure contiguity.

Overall, our approach takes advantage of the following MPI routines:
\begin{itemize}
\item \MPI{TYPE_CREATE_SUBARRAY}~\cite[p.~94]{mpistd31}. This routine
  constructs MPI datatypes describing an arbitrary non-strided slice
  of a dense multidimensional array. Subarray datatypes are routinely
  used in MPI-based codes and libraries to perform parallel MPI~I/O of
  distributed dense arrays, see~\cite[p.~207--208]{advmpi} for an
  executive example. A practical use of subarray datatypes and MPI~I/O
  can be found in PETSc, these features are used in the implementation
  of of parallel I/O for applications involving structured grids.
\item \MPI{ALLTOALLW}~\cite[p.~172]{mpistd31}. This routine is a
  generalized all-to-all scatter/gather collective communication
  operation allowing the specification of send and receive buffers
  with different datatypes, counts, and displacements for each process
  within an MPI communicator. To the best of our knowledge, this
  routine has not been widely used. A practical application we are
  aware of can be found in PETSc, where \MPI{ALLTOALLW} is used to
  implement scatter/gather operations on distributed vectors.
\end{itemize}
The use of these two routines can be illustrated with reference to
Fig.~\ref{fig:slab} and Fig.~\ref{fig:slab2D}.  First,
\MPI{TYPE_CREATE_SUBARRAY} is used to construct subarray datatypes
corresponding to the various array chuncks in Fig.~\ref{fig:slab2D:a}
and Fig.~\ref{fig:slab2D:c}. Afterwards, \MPI{ALLTOALLW} is fed with
these datatypes to perform the all-to-all exchange of array data from
the layout in Fig.~\ref{fig:slab:a} to the layout in
Fig.~\ref{fig:slab:b}. Thus, there is no need of the intermediate
remapping step depicted in Fig.~\ref{fig:slab2D:b}.

Alg.~\ref{alg:SubArray} and Alg.~\ref{alg:Exchange} show pseudocode
implementing the new global redistribution method, while
Listing~\ref{lst:SubArray} and Listing~\ref{lst:Exchange} present
corresponding implementations in the C programming language. Note that
these codes have no limitations on the dimensionality or arrays.

\begin{algorithm}[h!]
\caption{Subarray datatypes}
\label{alg:SubArray}
\begin{algorithmic}[1]
\Function{SubArray}{$T, N, v, M$}
\Input{$T$}{datatype}{elementary datatype descriptor}
\Input{$N$}{sequence}{local sizes of a $d$-dimensional array}
\Input{$v$}{integer}{axis to partition, $0 \leq v < d$}
\Input{$M$}{integer}{number of parts, $M > 0$}
\Output{$S$}{sequence}{subarray datatype descriptors}
\State $d \gets \mathbf{len}~N$
\For{$i \gets 0, d-1$}
\State $n(i) \gets N(i)$
\State $s(i) \gets 0$
\EndFor
\For{$p \gets 0, M-1$}
\State $n(v),\, s(v)\gets$ \Call{Decompose}{$N(v), M, p$}
\State $S(p) \gets$ \Call{CreateSubArray}{$T, N, n, s$}
\EndFor
\State \textbf{return} $S$
\EndFunction
\end{algorithmic}
\end{algorithm}

\begin{lstlisting}[float=h!,frame=single,
caption={Subarray datatypes},
label=lst:SubArray,
]
void subarray(MPI_Datatype datatype,
              int          ndims,
              int          sizes[ndims],
              int          axis,
              int          nparts,
              MPI_Datatype subarrays[nparts])
{
  int subsizes[ndims], substarts[ndims], n, s;
  for (int i = 0; i < ndims; i++)
    { subsizes[i] = sizes[i]; substarts[i] = 0; }
  for (int p = 0; p < nparts; p++) {
    decompose(sizes[axis], nparts, p, &n, &s);
    subsizes[axis] = n; substarts[axis] = s;
    MPI_Type_create_subarray(
        ndims, sizes, subsizes, substarts,
        MPI_ORDER_C, datatype, &subarrays[p]);
    MPI_Type_commit(&subarrays[p]);
  }
}
\end{lstlisting}

\begin{algorithm}[h!]
\caption{Exchange of arrays}
\label{alg:Exchange}
\begin{algorithmic}[1]
\Procedure{Exchange}{$P, A, v, B, w$}
\Input{$P$}{communicator}{group of communicating processes}
\Input{$A$}{array}{local array of elementary datatype $T$}
\Input{$v$}{integer}{axis of alignment for $A$}
\Output{$B$}{array}{local array of elementary datatype $T$}
\Input{$w$}{integer}{axis of alignment for $B$, $w \neq v$}
\State $T \gets$ \Call{Type}{$A$}\Comment{elementary datatype of array $A$}
\State $N_A \gets$ \Call{Shape}{$A$}\Comment{sequence with sizes of array $A$}
\State $N_B \gets$ \Call{Shape}{$B$}\Comment{sequence with sizes of array $B$}
\State $M \gets$ \Call{Size}{$P$}\Comment{number of processes in group}
\State $S_A \gets $ \Call{SubArray}{$T, N_A, v, M$}\Comment{sequence of datatypes for sending}
\State $S_B \gets $ \Call{SubArray}{$T, N_B, w, M$}\Comment{sequence of datatypes for receiving}
\State \Call{AllToAllW}{$P, A, S_A, B, S_B$}\Comment{generalized all-to-all scatter/gather}
\EndProcedure
\end{algorithmic}
\end{algorithm}

\begin{lstlisting}[float=h!,frame=single,
caption={Exchange of arrays},
label=lst:Exchange,
]
void exchange(MPI_Comm     comm,
              MPI_Datatype datatype,
              int          ndims,
              int          sizesA[ndims],
              void         *arrayA,
              int          axisA,
              int          sizesB[ndims],
              void         *arrayB,
              int          axisB)
{
  int nparts;
  MPI_Comm_size(comm, &nparts);
  MPI_Datatype subarraysA[nparts], subarraysB[nparts];
  subarray(datatype, ndims, sizesA, axisA, nparts, subarraysA);
  subarray(datatype, ndims, sizesB, axisB, nparts, subarraysB);
  int counts[nparts], displs[nparts];
  for (int p = 0; p < nparts; p++)
    { counts[p] = 1; displs[p] = 0; }
  MPI_Alltoallw(arrayA, counts, displs, subarraysA,
                arrayB, counts, displs, subarraysB, comm);
  for (int p = 0; p < nparts; p++) {
    MPI_Type_free(&subarraysA[p]);
    MPI_Type_free(&subarraysB[p]);
  }
}
\end{lstlisting}

In Alg.~\ref{alg:SubArray}, the \textsc{Decompose()} function from
Alg.~\ref{alg:Decompose} handles decompositions in $M$ parts by
computing the local lenghts and starting indices required to define
the output subarray datatype sequence $S=\{S_p\}_{p=0}^{M-1}$. Each
subarray datatype entry $S_p$ is created with calls to
\textsc{CreateSubArray()}, which represents an invocation to
\MPI{TYPE_CREATE_SUBARRAY}. The output datatype sequence $S$
effectively encodes a block-contiguous, one-dimensional partition in
$M$ chunks along a non-distributed axis of alignment $v$ for any
$d$-dimensional local array of elementary datatype $T$.

In Alg.~\ref{alg:Exchange}, the \textsc{SubArray()} function from
Alg.~\ref{alg:SubArray} is invoked to create subarray datatypes
sequences $S_A$ and $S_B$ using the shapes of the local input and
output arrays $A$ and $B$ and their respective axes of alignment $v$
and $w$.  Recalling Fig.~\ref{fig:slab}, $A$ corresponds to
source arrays with sizes $(N_0/P,N_1,N_2)$ as in
Fig.~\ref{fig:slab:a}, whereas $B$ corresponds to destination
arrays with sizes $(N_0,N_1/P,N_2)$ as in Fig.~\ref{fig:slab:b}. The
subarray datatype sequences $S_A$ and $S_B$ correspond to the various
chunks depicted in Fig.~\ref{fig:slab2D:a} and
Fig.~\ref{fig:slab2D:c}, respectively.  Finally, the call to
\textsc{AllToAllW()} represents an invocation of \MPI{ALLTOALLW} to
perform collective all-to-all exchange or array data. Evidently, there
is no need for local transposes or remappings.

Executing the call $\text{\textsc{Exchange}}(P, A, v, B, w)$ amounts
to a global redistribution within a process group $P$ from array $A$
in $v$-alignment to array $B$ in $w$-alignment, and in the previous
notation it corresponds to
\begin{equation}
B_{\ldots, j_w, \ldots, j_v/P, \ldots}
\xleftarrow[P]{v \rightarrow w}
{A}_{\ldots, j_w/P, \ldots, j_v, \ldots}.
\label{eq:gt}
\end{equation}
Note that the subarray datatypes created in Alg.~\ref{alg:Exchange}
and Listing~\ref{lst:Exchange} do not hold any array data in their
own; they are merely descriptors encoding array slicing operations.
The internal datatype handling engine within an MPI implementation is
able to decode the slicing information to complete the all-to-all
communication with the expected outcome, thus ensuring the black-box
nature of our approach. Consequently, rather than creating and
destroying datatypes as done in Listing~\ref{lst:Exchange}, a
production code should use Listing~\ref{lst:SubArray} in a setup phase
to create subarray datatypes, and reuse them as many times as needed
to perform data redistributions in one-line calls to \MPI{ALLTOALLW}.

\subsection{Cartesian process topologies}
\label{sec:cartesian}

Slab decompositions, as described in Sec.~\ref{sec:slab}, are
one-dimensional decompositions. Despite being very efficient in the
context of parallel FFTs, one-dimensional decompositions limit the
amount of parallelism that can be thrown at a problem. Recalling the
redistribution step in Eq.~(\ref{eq:gt_slab}), we necessarily require
$|P| \leq \text{min}(|j_0|,|k_1|)$. The next level of parallelism is
reached with multi-dimensional decompositions, to be presented shortly
in Sec.~\ref{sec:pencil}.  As a necessary prelude, this section will
discuss multi-dimensional Cartesian processor grids.

\begin{figure}[h!]
\centering
\begin{subfigure}[c]{0.5\textwidth}
  \centering
  \includegraphics[scale=0.5]{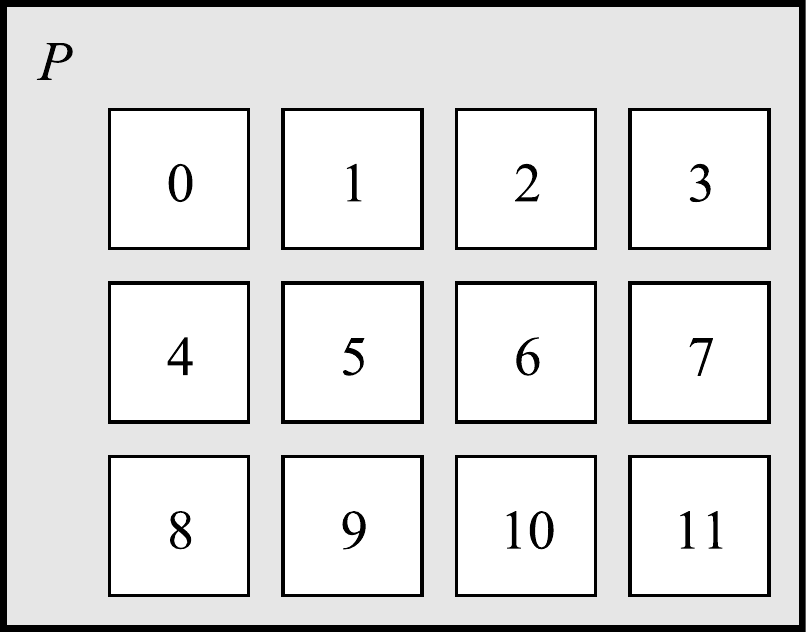}
  \caption{A process group $P$ with 12 processes.}
  \label{fig:subcomm:a}
\end{subfigure}%
\begin{subfigure}[c]{0.5\textwidth}
  \centering
  \includegraphics[scale=0.5]{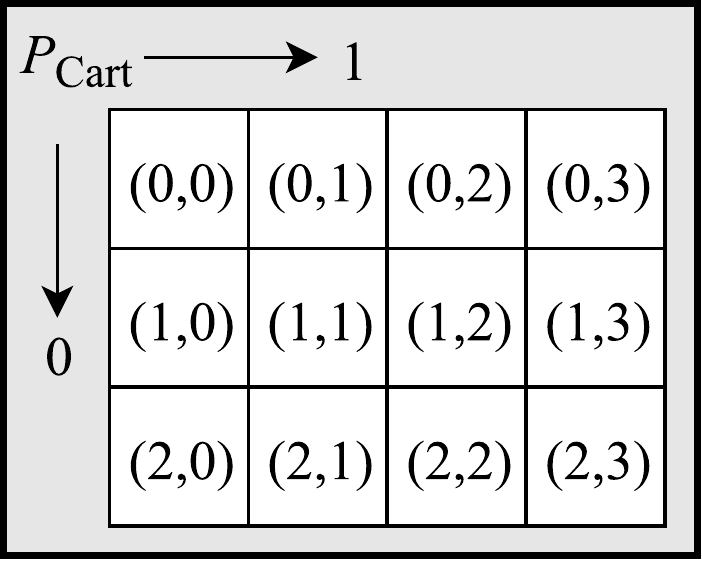}
  \caption{Cartesian grid with ${3}\times{4}$ processes.}
  \label{fig:subcomm:b}
\end{subfigure}\\
\begin{subfigure}[c]{0.5\textwidth}
  \centering
  \includegraphics[scale=0.5]{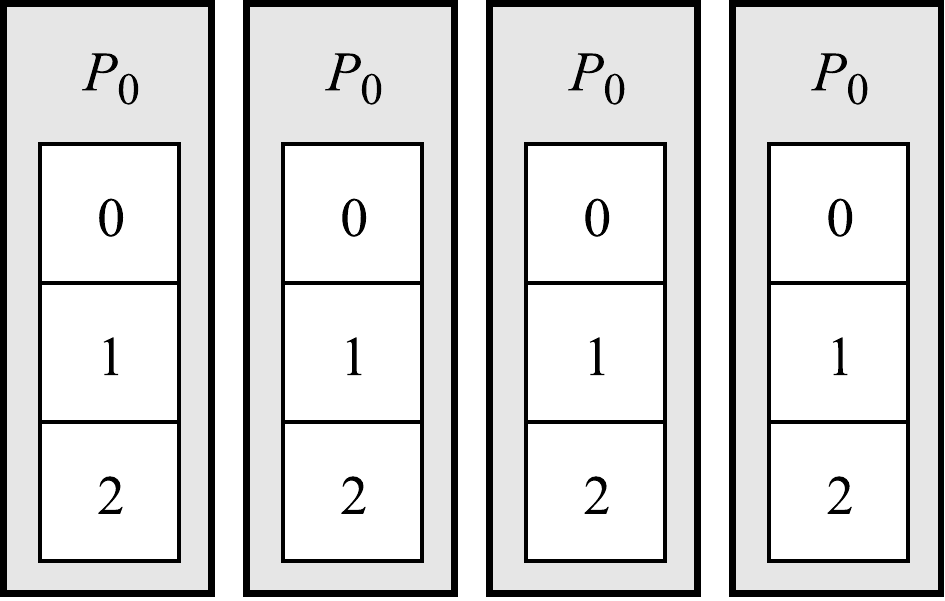}
  \caption{Subgroups $P_0$ in first direction.}
  \label{fig:subcomm:c}
\end{subfigure}%
\begin{subfigure}[c]{0.5\textwidth}
  \centering
  \includegraphics[scale=0.5]{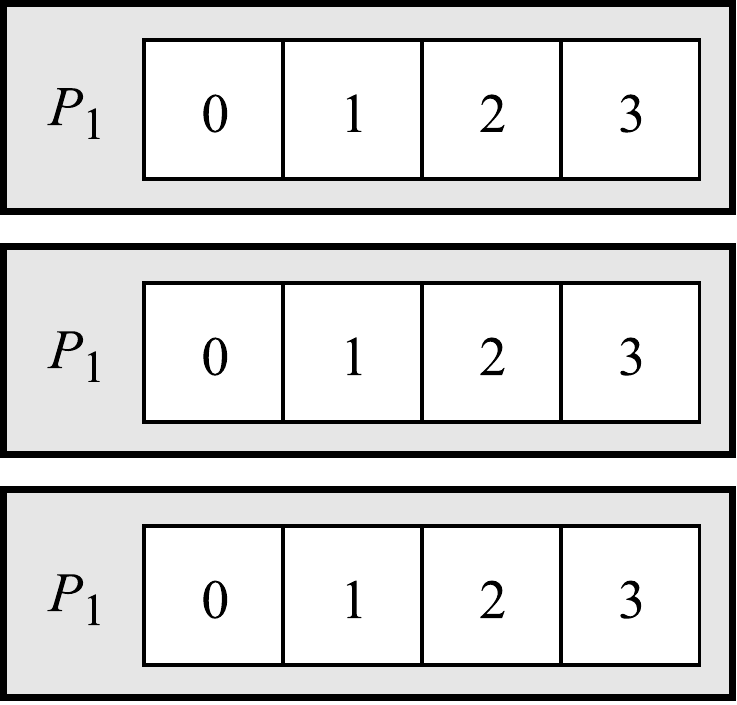}
  \caption{Subgroups $P_1$ in second direction.}
  \label{fig:subcomm:d}
\end{subfigure}
\caption{Two-dimensional Cartesian process grid and one-dimensional
  process subgroups corresponding to each direction.}
\label{fig:subcomm}
\end{figure}

Consider the rearrangement of a process group $P$ as a logically
two-dimensional Cartesian grid of $M_0 \times M_1$ processes such that
$|P| = M_0 \cdot M_1$. For each process in $|P|$ with identifier
$p=0,1,\ldots,|P|-1$ we assign a two-tuple of process coordinates
$(p_0, p_1)$ with $p_i=0,1,\ldots,M_i-1$ corresponding to each
direction $i=0,1$. Such assignment of coordinates induces a
partitioning of the Cartesian topology in one-dimensional subgroups
corresponding to each direction. In the first direction, we obtain
$M_1$ subgroups collectively denoted $P_0$, each with $|P_0| = M_0$
processes with identifiers $p_0$. Similarly, in the second direction,
we obtain $M_0$ subgroups $P_1$ collectively denoted $P_1$, each with
$|P_1| = M_1$ processes with identifiers $p_1$. Fig.~\ref{fig:subcomm}
depicts these steps for a group of $12$ processes arranged as a
two-dimensional grid of ${3}\times{4}$ processes. The generalization
to higher dimensions is straightforward.

The MPI standard provides many facilities for managing Cartesian
processor grids. The utility routine
\MPI{DIMS_CREATE}~\cite[p. 293]{mpistd31} can be used to compute a
balanced distribution of processes among a given number of
dimensions. The routine \MPI{CART_CREATE}~\cite[p. 292]{mpistd31}
constructs process groups with attached Cartesian topology. Finally,
the routine \MPI{CART_SUB}~\cite[p. 311]{mpistd31} partitions a
Cartesian topology in lower-dimensional subgroups. These routines are
combined in Listing~\ref{lst:SubComm} to define a Cartesian topology
and obtain the partitions corresponding to each direction. Note that
this code can handle processor grids of any dimensionality.

\begin{lstlisting}[float=h!,frame=single,%
caption={One-dimensional subgroups of a Cartesian process grid},%
label=lst:SubComm,%
]
void subcomm(MPI_Comm comm,
             int      ndims,
             MPI_Comm subcomms[ndims])
{
  MPI_Comm comm_cart;
  int nprocs, dims[ndims], periods[ndims], remdims[ndims];
  for (int i = 0; i < ndims; i++)
    { dims[i] = periods[i] = remdims[i] = 0; }
  MPI_Comm_size(comm, &nprocs);
  MPI_Dims_create(nprocs, ndims, dims);
  MPI_Cart_create(comm, ndims, dims, periods, 1, &comm_cart);
  for (int i = 0; i < ndims; i++) {
    remdims[i] = 1;
    MPI_Cart_sub(comm_cart, remdims, &subcomms[i]);
    remdims[i] = 0;
  }
  MPI_Comm_free(&comm_cart);
}
\end{lstlisting}

\subsection{Pencil decomposition}
\label{sec:pencil}

A pencil decomposition makes use of two subgroups of processors, and
distributes two index sets simultaneously in a multidimensional array.
A parallel FFT on a three-dimensional array that is initially
distributed with processor groups $P_0$ and $P_1$, can be performed in
five steps:
\begin{align}
\tilde{u}_{j_0/P_0, j_1/P_1,  k_2} &= \mathcal{F}_2\left( u_{j_0/P_0, j_1/P_1, j_2} \right), \label{eq:pencil0} \\
\tilde{u}_{j_0/P_0, j_1, k_2/P_1} &\xleftarrow[P_1]{2\rightarrow 1} \tilde{u}_{j_0/P_0, j_1/P_1, k_2} , \label{eq:gtp1} \\
\tilde{u}_{j_0/P_0, k_1, k_2/P_1} &= \mathcal{F}_1(\tilde{u}_{j_0/P_0, j_1, k_2/P_1}), \label{eq:pencil1} \\
\tilde{u}_{j_0, k_1/P_0, k_2/P_1} &\xleftarrow[P_0]{1\rightarrow 0} \tilde{u}_{j_0/P_0, k_1, k_2/P_1} , \label{eq:gtp2} \\
\hat{u}_{k_0, k_1/P_0, k_2/P_1} &= \mathcal{F}_0(\tilde{u}_{j_0, k_1/P_0, k_2/P_1}).\label{eq:pencil2}
\end{align}
For a higher-dimensional array the procedure is exactly the same,
except that the initial array in Eq.~(\ref{eq:pencil0}) is partially
transformed over more trailing axes, like in Eq.~(\ref{eq:slab0}).

\begin{figure}[h!]
  \centering
  \begin{subfigure}[t]{0.30\textwidth}
    \centering
    \includegraphics[width=\textwidth]{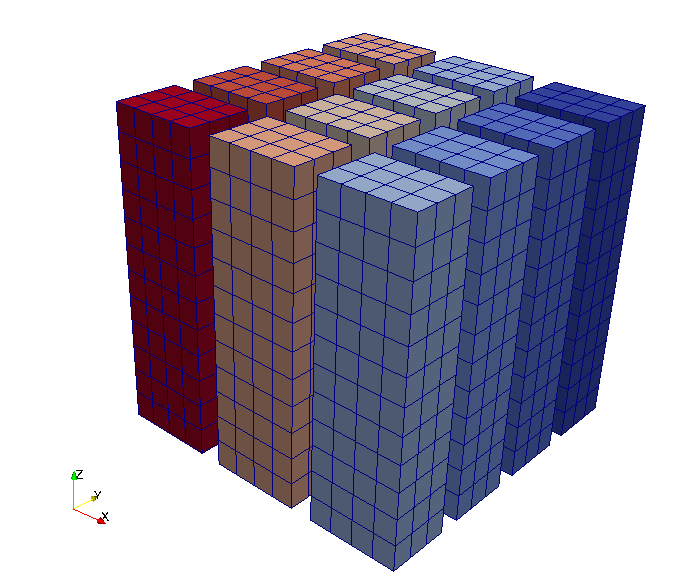}
    \caption{Pencils in 2D decomposition aligned in $z$-direction.}
    \label{fig:pencil:a}
  \end{subfigure}
  \hfill
  \begin{subfigure}[t]{0.30\textwidth}
    \centering
    \includegraphics[width=\textwidth]{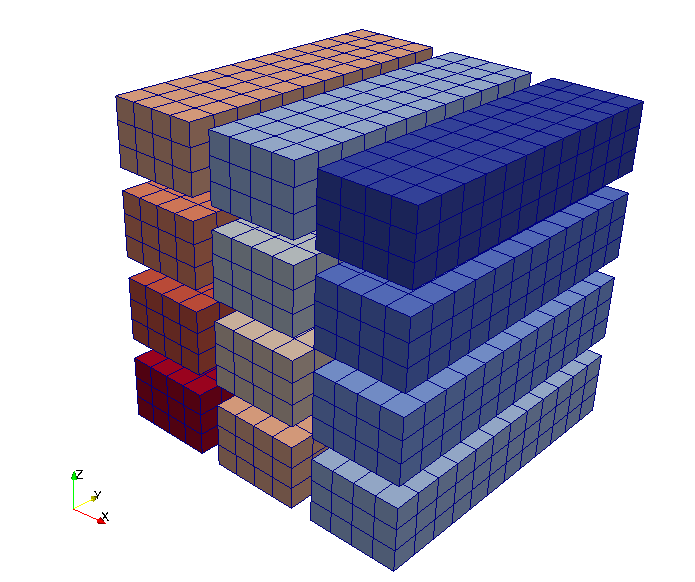}
    \caption{Pencils in 2D decomposition aligned in $y$-direction.}
    \label{fig:pencil:b}
  \end{subfigure}
  \hfill
  \begin{subfigure}[t]{0.31\textwidth}
    \centering
    \includegraphics[width=\textwidth]{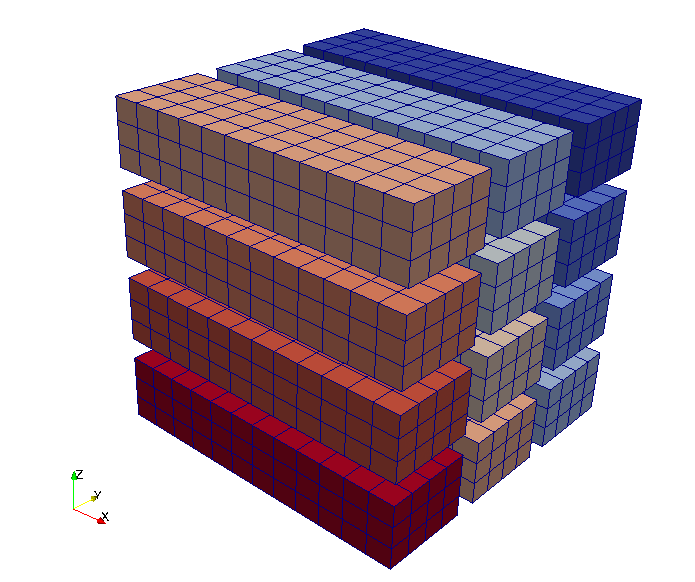}
    \caption{Pencils in 2D decomposition aligned in $x$-direction.}
    \label{fig:pencil:c}
  \end{subfigure}
  \caption{2D pencil decomposition on a $3 \times 4$ process grid for
    three different alignments of a global 3D array.  Each local
    subarray (or \emph{pencil}) is colored in correspondence to the
    owning process, from red (process $0$) to blue (process $11$).}
  \label{fig:pencil}
\end{figure}

To illustrate the procedure, Fig.~\ref{fig:pencil} shows a 3D array of
global size $12^3$ decomposed on a $3 \times 4$ Cartesian process
grid. Each local array, or pencil, is colored to identify the owning
process. In reference to Fig.~\ref{fig:subcomm}, the deep-red
pencils are owned by process $(0,0)$ in $P_\text{Cart}$, which
corresponds to process $0$ in group $P$. Similarly, the deep-blue
pencils are owned by process $(2,3)$ in $P_\text{Cart}$, which
corresponds to process $11$ in group $P$. The global array in
Fig.~\ref{fig:pencil:a} is initially aligned in axis $2$
($z$-direction), where the index sets $j_0$ and $j_1$ are distributed
on the four subgroups $P_0$ and the three subgroups $P_1$ (see
Fig.~\ref{fig:subcomm:c} and Fig.~\ref{fig:subcomm:d}), respectively.
After the partial transform (Eq.~\ref{eq:pencil0}) over axis $2$, a
global redistribution (Eq.~\ref{eq:gtp1}) is performed to realign the
global array in axis $1$ ($y$-direction). Fig.~\ref{fig:pencil:b}
shows this intermediate alignment, where index sets $j_0$ and $k_2$ are
distributed on subgroups $P_0$ and $P_1$, respectively. An additional
partial transform on axis $1$ (Eq.~\ref{eq:pencil1}) and global
redistribution (Eq.~\ref{eq:gtp2}) lays the array in its final
alignment in axis $0$ ($y$-direction) as shown in
Fig.~\ref{fig:pencil:c}, where index sets $k_1$ and $k_2$ are
distributed on subgroups $P_0$ and $P_1$, respectively. Finally, a
partial transform on axis $0$ (Eq.~\ref{eq:pencil2}) completes the
procedure.

\begin{figure}[h!]
  \centering
  \begin{subfigure}[t]{0.30\textwidth}
    \centering
    \includegraphics[width=\textwidth]{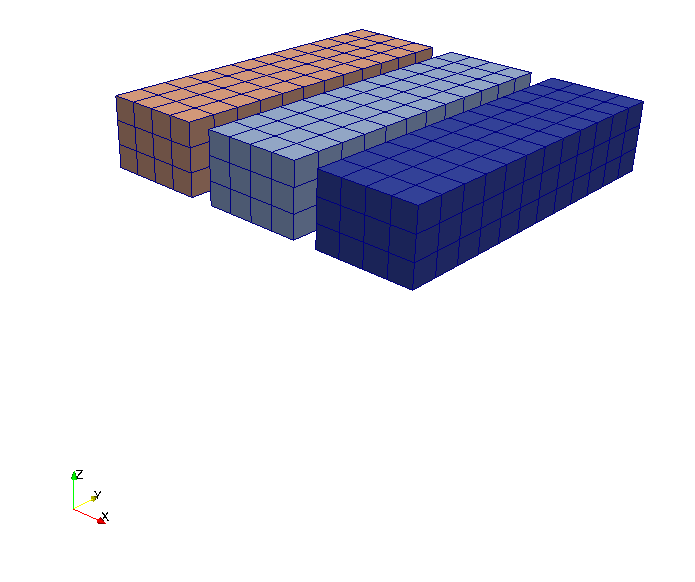}
    \caption{Top subslab of Fig.~\ref{fig:pencil:b}.}
  \end{subfigure}
  \hskip 0.1\textwidth
  \begin{subfigure}[t]{0.30\textwidth}
    \centering
    \includegraphics[width=\textwidth]{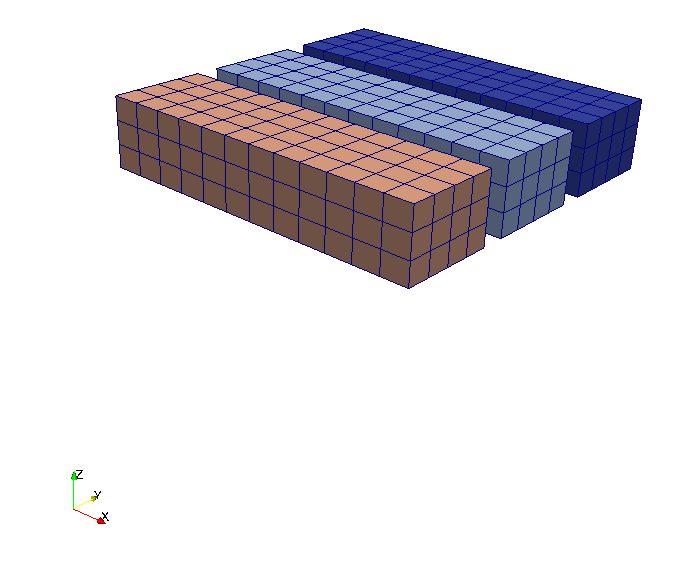}
    \caption{Top subslab of Fig.~\ref{fig:pencil:c}.}
  \end{subfigure}
  \caption{Subslabs within a 2D pencil decomposition, corresponding to
    one of the subgroups $P_{0}$ with process $\{3, 7, 11\}$, before
    (a) and after (b) a global redistribution from $y$-alignment to
    $x$-alignment.}
  \label{fig:pencil_as_slab}
\end{figure}

At first glance, the pencil method in
Eqs.~(\ref{eq:pencil0}--\ref{eq:pencil2}) looks substantially more
complex to implement than the slab method in
Eqs.~(\ref{eq:slab0}--\ref{eq:slab1}). This is indeed the case in
traditional implementations following the approach of
Sec.~\ref{sec:old_redist}. Codes typically accumulate hundreds of
lines with tedious nested loops just to implement local remappings for
the various possible alignments. Furthermore, these pieces of code are
usually hardwired to work in the three-dimensional case, and
generalizations to higher dimensions are a daunting task. %
At this point, with the support of Fig.~\ref{fig:pencil_as_slab}, we
make a key although simple observation: %
\emph{a 2D pencil decomposition can be reinterpreted as a collection
  of slab decompositions on one-dimensional process subgroups of the
  two-dimensional process grid}. %
The consequence is remarkable: global redistribution operations like
the ones in Eqs.~(\ref{eq:gtp1}) and (\ref{eq:gtp2}) can be performed
just by concurrently executing Alg.~\ref{alg:Exchange} on subslabs
within each process subgroup. After constructing one-dimensional
process subgroups as in Listing~\ref{lst:SubComm}, the pseudocodes and
listings presented in Sec.~\ref{sec:new_redist} can be reused
\emph{verbatim} to perform the two global redistribution steps
required in the 2D pencil method\footnote{Note that PFFT uses similar
ideas to implement multidimensional transforms based on the slab
code available in FFTW.}. See Appendix~\ref{app:3D} for a C code
listing showcasing full forward and backward complex-to-complex
transforms of a three-dimensional array with 2D pencil decomposition.

\subsection{Higher-dimensional decompositions}
\label{sec:4d}

A $d$-dimensional array can be distributed on at most
$(d-1)$-dimensional process grids, such that an initial partial FFT
can be performed in at least one non-distributed direction. Subsequent
global redistributions and partial transforms follow to complete a
full multidimensional parallel FFT. The reinterpretation of pencil
decompositions as collections of slab decompositions generalizes to
higher-dimensional arrays and process grids. Once again, the
pseudocodes and listings presented in Secs.~\ref{sec:slab}
and~\ref{sec:cartesian} can be reused verbatim to perform any global
redistribution step.

As a proof of concept, consider a parallel FFT of a four-dimensional
array, $u_{j_0, j_1, j_2, j_3}$, using a three-dimensional process
grid decomposed in the various one-dimensional process subgroups
$P_0$, $P_1$ and $P_2$. These process subgroups can be generated with
Listing~\ref{lst:SubComm}. We perform a parallel FFT on such an array
in seven steps (four partial transforms and three global
redistributions\footnote{In general, a $d$-dimensional array
distributed on a $(d-1)$-dimensional processor grid requires $d$
partial Fourier transform and $d-1$ global redistributions steps.})
as follows:
\begin{align}
\tilde{u}_{j_0/P_0, j_1/P_1,  j_2/P_2,k_3} &= \mathcal{F}_3\left( u_{j_0/P_0,
j_1/P_1, j_2/P_2, j_3} \right), \label{eq:4d_fft0}\\
\tilde{u}_{j_0/P_0, j_1/P_1, j_2,k_3/P_2} &\xleftarrow[P_2]{3\rightarrow 2} \tilde{u}_{j_0/P_0, j_1/P_1, j_2/P_2,k_3}, \label{eq:4dgtp1} \\
\tilde{u}_{j_0/P_0, j_1/P_1, k_2, k_3/P_2} &= \mathcal{F}_2(\tilde{u}_{j_0/P_0,
j_1/P_1, j_2, k_3/P_2}), \label{eq:4d_fft1}\\
\tilde{u}_{j_0/P_0, j_1, k_2/P_1, k_3/P_2} &\xleftarrow[P_1]{2\rightarrow 1} \tilde{u}_{j_0/P_0, j_1/P_1, k_2, k_3/P_2}, \label{eq:4dgtp2} \\
\tilde{u}_{j_0/P_0, k_1, k_2/P_1, k_3/P_2} &= \mathcal{F}_1(\tilde{u}_{j_0/P_0,
j_1, k_2/P_1, k_3/P_2}), \label{eq:4d_fft2} \\
\tilde{u}_{j_0, k_1/P_0, k_2/P_1, k_3/P_2} &\xleftarrow[P_0]{1\rightarrow 0} \tilde{u}_{j_0/P_0, k_1, k_2/P_1, k_3/P_2}, \label{eq:4dgtp3} \\
\hat{u}_{k_0, k_1/P_0, k_2/P_1, k_3/P_2} &= \mathcal{F}_0(\tilde{u}_{j_0,
k_1/P_0, k_2/P_1, k_3/P_2}).\label{eq:4d_fft3}
\end{align}
The global redistribution steps in Eqs.~(\ref{eq:4dgtp1}),
(\ref{eq:4dgtp2}) and (\ref{eq:4dgtp3}) can be performed with
concurrent executions of Alg.~\ref{alg:Exchange} on the proper
subslabs corresponding to each process subgroup. See
Appendix~\ref{app:4D} for a C code listing showcasing full forward and
backward complex-to-complex transforms of a four-dimensional array
with three-dimensional decomposition.

Note that all intermediate arrays in
Eqs.~(\ref{eq:4d_fft0}--\ref{eq:4d_fft3}) must be preallocated before
the 4D parallel FFT can be executed, since the global redistribution
steps are out-of-place. In Appendix~\ref{app:4D} four arrays are
preallocated since there are four differently shaped arrays in
(\ref{eq:4d_fft0}--\ref{eq:4d_fft3}) for a complex-to-complex
transform.  This may seem like excessive use of memory. However, since
all intermediate arrays can be allocated in contiguous memory, the
method could, in practice, simply use the two largest intermediate
arrays as work buffers for all intermediate steps, regardless of
dimensionality.

\section{Performance evaluation}
\label{sec:results}

We will now explore the efficiency of the new global redistribution
method proposed in previous sections. We first want to remind the
reader that what is proposed is really a black-box method applicable
to any array dimensionality and processor mesh decomposition. The
executable code required is shown to be approximately 50 lines of code
in C.  Considering the complexity normally associated with this task
(local transposes with or without changes of strides, in-place or
out-of-place), and the thousands of lines of code dedicated to global
redistribution by other parallel FFT vendors, there is at the outset
of this section something to be said for simplicity.

With the current proposed method the global redistribution is achieved
in one single call to \MPI{ALLTOALLW}. The cost of this call should be
compared to the entire global redistribution operation implemented by
other packages, that are typically using \MPI{ALLTOALL(V)} merely for
communication. Now, \MPI{ALLTOALL(V)} works on contiguous dataarrays
in both ends, both for sending and receiving ranks, and there are
highly optimised versions available on several architectures.  On the
contrary, the subarray type used by \MPI{ALLTOALLW} is in general
discontiguous, and there are to the authors' knowledge no
architecture-specific optimizations available. This represents a
significant disadvantage of the current proposed method.  However, if
the additional cost of the non-optimized \MPI{ALLTOALLW} is not higher
than the time spent on local remappings by other global redistribution
methods, then it can still be competitive.  Note that the collective
communication routines have several different implementations by
different vendors. MPICH, for example, has four different
implementations of \MPI{ALLTOALL} that are called based on the size of
the involved arrays. For \MPI{ALLTOALLW}, on the other hand, a
non-blocking \MPI{ISEND}/\MPI{IRECV} algorithm is used regardless the
array size.

Computations are performed on the Shaheen Cray XC40 supercomputer,
with its primary resource capable of 7.2 Petaflops peak (5.5 sustained
on the HPL benchmark \cite{top500}).  The computer comes with highly
optimised, preinstalled versions of the MPICH and FFTW libraries, and
we use these libraries for all codes. Furthermore, all codes are
compiled with Cray compilers using similar compiler options, and
multithreading is disabled. The Cray XC system has 6,174 dual sockets
compute nodes based on 16-core Intel Haswell processors running at
2.3GHz. Each node has 128GB of DDR4 memory running at 2.3GHz. With
this multicore hardware technology there are two very different
communication speeds at play - the shared intra-node and the
distributed inter-node communication. Within each node there are
possibly 32 cores that communicate with each other using a shared
in-node memory, whereas across nodes the communication uses the Cray
Aries interconnect with Dragonfly topology, which requires at most
three hops between any two cores globally. The Cray XC comes with
several architecture-specific optimizations for
\MPI{ALLTOALL(V)}. These optimizations may be turned off using
environment variable \texttt{MPICH\_COLL\_OPT\_OFF}, in which case
they will use the same non-blocking \MPI{ISEND}/\MPI{IRECV} as
\MPI{ALLTOALLW}. This has not been done here.

The global redistribution method described in previous chapters only
directly affects the parallel FFT in steps (\ref{eq:gt_slab},
\ref{eq:gtp1}, \ref{eq:gtp2}). Furthermore, the sequential FFTs can be
performed using any FFT vendor, and we can mainly affect the
efficiency by carefully obtaining optimised installations on any given
platform. However, some global/local transpose operations performed by
other codes are done with the purpose of speeding up the sequential
FFT, e.g., by aligning data contiguously in memory before
executing. Hence, we will not only look at the global redistribution,
but rather the complete transform. To this end we have implemented
both slab and pencil 3D codes in C, using the global redistribution
code from Sec.~\ref{sec:mpi}. Apart from this, the implementation is
quite trivial, just a matter of creating processor groups, allocating
arrays of the correct shapes and planning serial FFTs. For
completeness, a 3D pencil code for complex-to-complex transforms is
shown in Appendix \ref{app:3D}.  We compare our code with P3DFFT, FFTW
and 2DECOMP\&FFT, where FFTW only has the slab method implemented, and
the other two are primarily advertised as pencil decomposition
codes. For the parallel FFTW results, with slab decomposition, we have
used \texttt{fftw\_mpi\_plan\_dft\_r2c\_3d} and
\texttt{fftw\_mpi\_plan\_dft\_c2r\_3d} with the transposed out option.
2DECOMP\&FFT has been compiled with the C preprocessor flag
\texttt{-DOVERWRITE}.  P3DFFT has been compiled using configure
options \texttt{./configure --enable-cray --enable-fftw
  --enable-measure FC=ftn CC=cc}.  We have compiled both with and
without the \emph{stride1} option for the global redistribution, but,
since without has been found to be generally faster, only the results
of the code that disables the \emph{stride1} option are reproduced
here.  We use \texttt{FFTW\_MEASURE} for all codes in planning FFTs,
and the pencil decomposition is as chosen by the \MPI{DIMS_CREATE}
function.  Since P3DFFT and 2DECOMP\&FFT are implemented in
column-major Fortran, we here use arrays of transposed dimensions as
compared to the C codes.  That is, when using a global array of shape
$(N_0, N_1, N_2)$ in C, then an array of shape $(N_2, N_1, N_0)$ is
used correspondingly in Fortran.  Furthermore, Fortran codes use a
processor mesh that has been transposed compared to the C codes.  For
all codes we compute performance using two nested loops. The inner
loop performs 3 consecutive, uninterrupted, forward/backward
transforms, and this inner loop is repeated 50 times in an outer loop,
with an MPI barrier call at the outset. The measured time after the
three inner loops is reduced to the maximum value across all
processors. From these values we then choose to report the fastest
result from the 50 outer loops, divided by 3.  For our C-code, P3DFFT
and 2DECOMP\&FFT we also place timers around each one of the major
steps involved. This is not done for FFTW, because it would require a
recompilation of the code, and we want to make use of the optimised
Cray version.

\begin{figure}
\includegraphics[width=\textwidth]{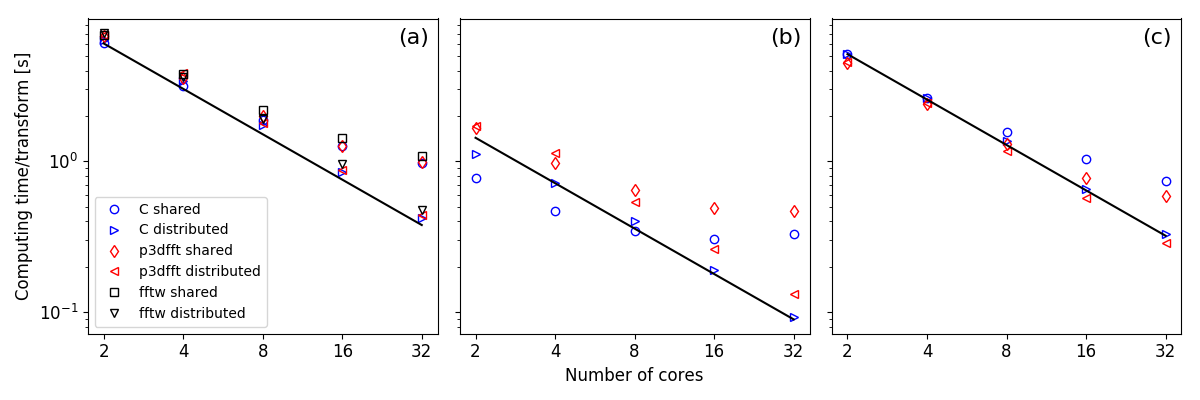}
\caption{Strong scaling with slab decomposition of complete
real-to-complex/complex-to-real FFTs on a mesh that has global shape
$700^3 $.  Showing fastest times measured. The shared results
correspond to shared intra-node operation, whereas distributed refer
to one core per node operation. Subplot (a) shows total time,
whereas (b) and (c) shows times for global redistribution and FFTs,
respectively.
\label{fig:Strong_slab_700}}
\end{figure}

The first results are computed with FFT codes that come with dedicated
slab implementations, using only up to 32 processor cores, such that
the two different modes of communication (intra-node vs inter-node)
can be easily compared. One set of simulations employs one CPU core
per node, and the other employs cores from one single node. The two
settings are referred to as distributed and shared simulations,
respectively. We use a quite large double precision input array of
global shape $700^3$. Figure \ref{fig:Strong_slab_700} a) shows the
strong scaling achieved by our C-code, FFTW and P3DFFT (compiled with
option \emph{oned} enabled), for both distributed and shared
operations. It is evident that all codes behave somewhat similarly,
showing good strong scaling in the purely distributed mode, whereas
the scaling is poor for purely shared mode. For all numbers of cores
our C-code is fastest, followed by P3DFFT and FFTW. For a more
detailed inspection, Figure \ref{fig:Strong_slab_700} b) and c) show
the individual timings for P3DFFT and our C-code for global
redistributions and serial FFTs, respectively. We see that P3DFFT
achieves somewhat faster serial FFTs, but that a larger difference is
seen for the global redistributions in b), where our method is
significantly faster over the entire range of cores. Again, scaling is
seen to be good only for the purely distributed inter-node mode of
operation.  A deeper inspection, using Cray's perftools, reveals that
the inter-node operation allows for significantly higher clock speeds,
with frequencies up to 3.5GHz, compared to the intra-node operation
that clocks in around 2.5GHz for the highest number of cores. This
slow-down explains the poor scaling of the serial FFTs measured in
Fig. \ref{fig:Strong_slab_700} c).  The poor performance of the shared
intra-node mode of operation is well known and has been the center of
much focus, especially for supercomputers, which have been moving
towards multicore designs, see, e.g., Kumar et al. \cite{kumar08}.
Consequently, MPICH comes with some relevant compiler settings,
especially for \MPI{ALLTOALL} (e.g.,
\texttt{MPICH\_SHARED\_MEM\_COLL\_OPT}), which enables code that tries
to take advantage of the shared memory.  2DECOMP\&FFT has an
implementation tailored to take advantage of the shared memory, to
limit the number of messages being sent (so-called leader based
algorithm). However, we have not made use of such multicore aware
algorithms here, and 2DECOMP\&FFT has been used without the shared
memory option. Furthermore, we have only used MPICH with default
settings.

\begin{figure}
\includegraphics[width=\textwidth]{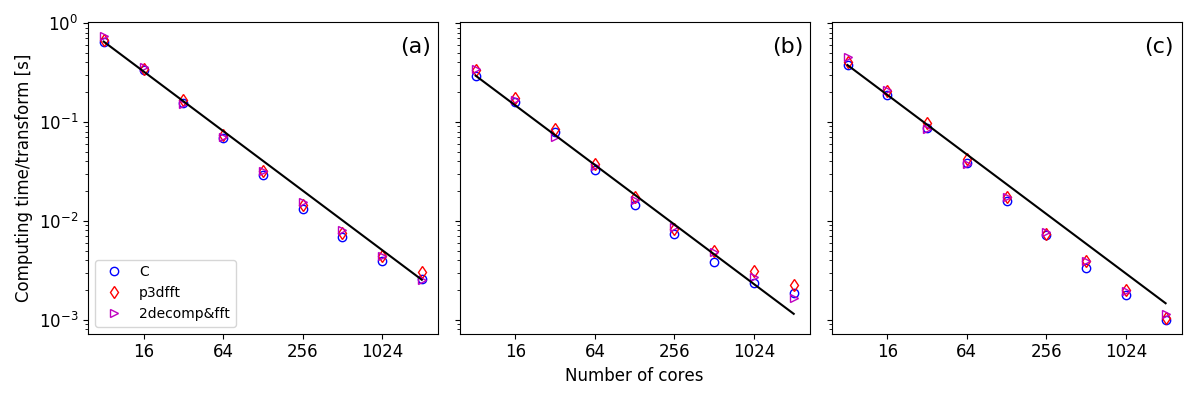}
\caption{Strong scaling with pencil decomposition of complete
real-to-complex/complex-to-real FFTs on a mesh that has global shape
$512^3 $. Showing fastest times measured. Subplot (a) shows total
time, whereas (b) and (c) shows times for global redistribution and
FFTs, respectively.
\label{fig:Strong_pencil_complete}}
\end{figure}

We next consider the pencil decomposition in purely distributed
inter-node mode and perform a strong scaling study of forward and
backward transforms on a global array of double precision and size
$512^3$ in physical space.  Figure \ref{fig:Strong_pencil_complete}
shows the fastest measured times for complete forward and backward
transforms of our C-code, P3DFFT and 2DECOMP\&FFT, for a wide scaling
range. Throughout the entire range our C-code is found in a) to be
5-10 \% faster than P3DFFT and 1-5 \% faster than 2DECOMP\&FFT.  For
all codes the scaling is more than excellent, achieving optimal speed
per core at 256 process cores, corresponding to a mesh of size 524,288
($64^2\cdot 128$) per core.  Figure~\ref{fig:Strong_pencil_complete}
b) and c) shows the corresponding individual timings for global
redistributions and FFTs.  For P3DFFT and 2DECOMP\&FFT the timings for
global redistributions are simply computed as the time it takes for
one (forward or backward) transform minus the time spent inside
sequential FFTs. It is evident from c) that the advantage of our
C-code for this case is obtained through faster global
redistributions, and there is little difference in computing times for
sequential FFTs. The apparent superunitary scaling is explained by the
higher frequencies achieved by the processors at larger core counts.

\begin{figure}
\centering
\includegraphics[width=\textwidth]{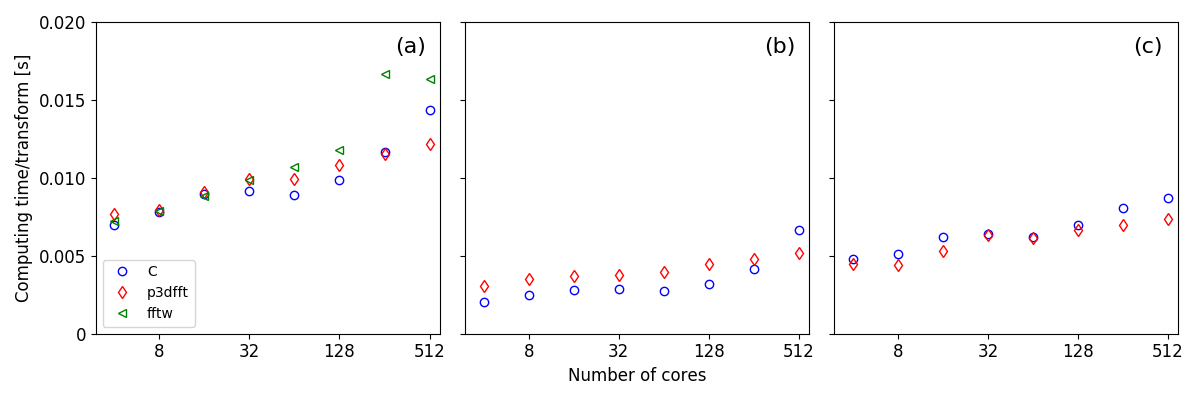}
\caption{Weak scaling of slab decomposition of complete
real-to-complex/complex-to-real FFT on a mesh that is $524,288$ per
processor. Showing fastest time measured.  Subplot (a) shows total
time, whereas (b) and (c) shows times for global redistribution and
FFTs, respectively.
\label{fig:Weak_slab_complete}}
\end{figure}

Next we perform a weak scaling study, using local arrays of double
precision and a size (524,288) corresponding to a grid of shape
$64^2\cdot128$ in real space, because in the previous strong test
(Fig. \ref{fig:Strong_pencil_complete}) the Shaheen computer was found
to be most efficient (speed per core) for arrays of this size. We
consider first the slab decomposition, and compare to the fastest
results obtained with FFTW and P3DFFT in Figure
\ref{fig:Weak_slab_complete}.  We observe in a) that the new C-code is
equally fast as FFTW for small processor counts (4, 8, 16), and faster
for higher (32 to 512). P3DFFT is fastest for the highest number of
cores, where each slab is only one layer thick, but generally 5-10\%
slower the other runs. FFTW scales poorly at higher than 128 cores,
but it is apparent that also the new method scales rather poorly in
this limit of very thin slabs, i.e., when we approach the maximum
number of cores possible for the slab method. P3DFFT is showing the
best scaling in this limit. The major reason for the faster results
obtained with the new method on low counts is found when we isolate
the global redistributions.  The computing times spent on global (and
local) redistributions (i.e., computing time outside sequential FFTs)
are shown in Figure \ref{fig:Weak_slab_complete} b).  Again it is
evident that the new method is not highly efficient in the limit of
very this slabs, but all over, the global redistributions are
approximately 40-50\% faster than for P3DFFT.  To complete the
picture, we also plot the fastest computing time for the sequential
FFTs in Fig. \ref{fig:Weak_slab_complete} c). Here it is evident that
some of the sequential transforms are faster with P3DFFT, possibly
because of the alignment of intermediate arrays. But the faster serial
FFTs are not sufficiently much better that they can close the gap
introduced by the global redistributions.

\begin{figure}
\centering
\includegraphics[width=\textwidth]{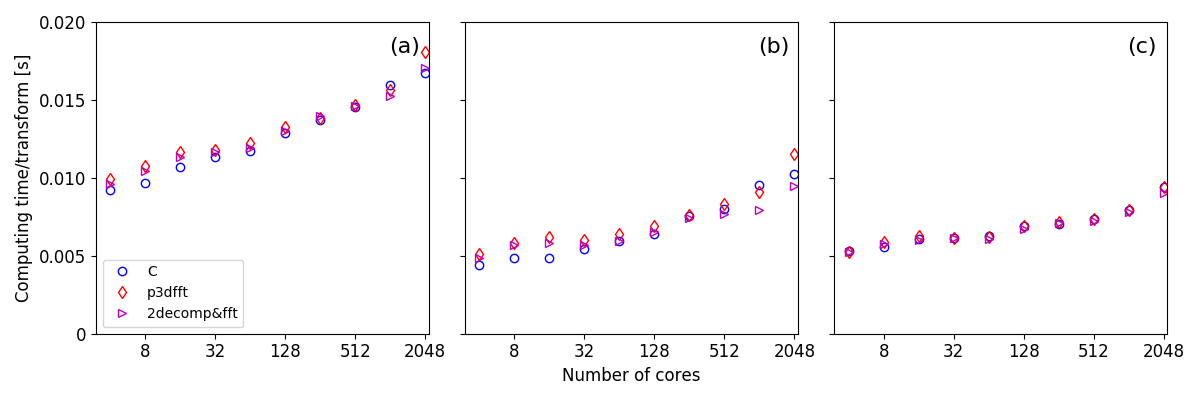}
\caption{Weak scaling of pencil decomposition for complete
real-to-complex/complex-to-real FFT on a mesh that is size $524,288$
per processor. Showing fastest times measured. Subplot (a) shows
total time, whereas (b) and (c) shows times for global
redistribution and FFTs, respectively.
\label{fig:Weak_complete}}
\end{figure}

We now perform the same weak efficiency test for the pencil
decomposition.  Figure \ref{fig:Weak_complete} shows the fastest
recorded times for a complete forward and backward transform. We see
that our method is slightly faster than both P3DFFT and 2DECOMP\&FFT
for most measured core counts, but that differences are small.
Breaking it down further, Fig. \ref{fig:Weak_complete} c) shows the
weak scaling (fastest recorded over 50 outer loops, 3 inner) for the
sum of the 6 sequential FFTs required to do the forward and backward
transforms.  We note that there is hardly any difference at all
between the codes over the entire range of the study.  Figure
\ref{fig:Weak_complete} b) shows the total cost of global
redistribution steps (\ref{eq:gtp1}) and (\ref{eq:gtp2}) for one
forward and one backward transform. As for the total transform,
Fig. \ref{fig:Weak_complete} a), our C-code is fastest for low core
counts, whereas there is less separating the codes for core counts
larger than 128.

\begin{figure}
\centering
\includegraphics[width=\textwidth]{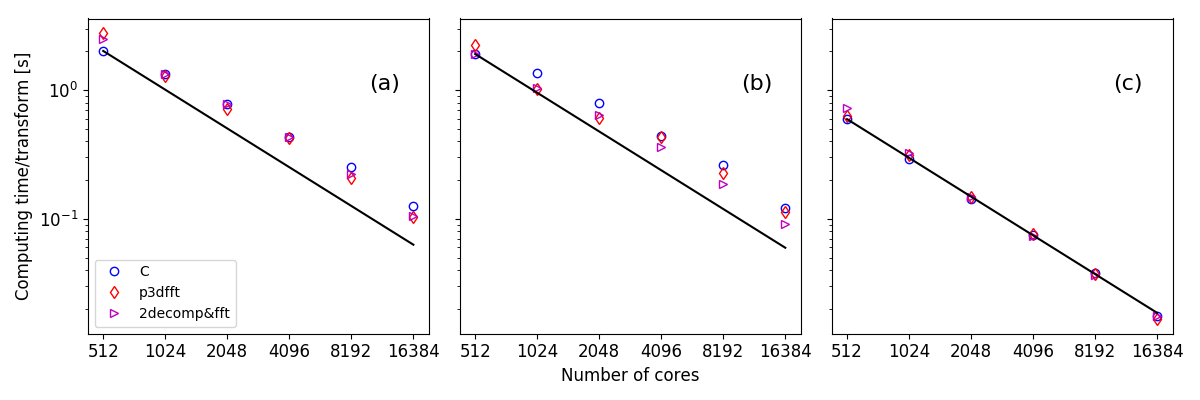}
\caption{Strong scaling of pencil decomposition for complete
real-to-complex/complex-to-real FFT on a mesh with global shape
$2048^3$. Using 16 cores per node for mixed multicore (inter- and
intra-node) communication. Showing fastest times measured. Subplot
(a) shows total time, whereas (b) and (c) shows times for global
redistribution and FFTs, respectively.
\label{fig:Strong_pencil_2048}}
\end{figure}

Figures \ref{fig:Strong_pencil_complete}, \ref{fig:Weak_slab_complete}
and \ref{fig:Weak_complete} have been generated in the fully
distributed, inter-node mode of operation. This mode is the fastest
per core, but for supercomputers one normally has to pay CPU-hours for
the entire node, even if only one core is used per node. For this
reason it is also important to investigate the performance of the
mixed multicore (inter- and intra-node) communication mode. To this
end Fig. \ref{fig:Strong_pencil_2048} shows the strong scaling of our
C-code, P3DFFT and 2DECOMP\&FFT for a global mesh of shape $2048^3$ in
real space, using 16 cores per node. Here it is evident that the
\MPI{ALLTOALL(V)} based global redistribution is faster, at least when
the mesh per node is large. There is less separating the methods as
the number of cores increases, where there is less work on each node.

\begin{figure}
\centering
\includegraphics[width=0.6\textwidth]{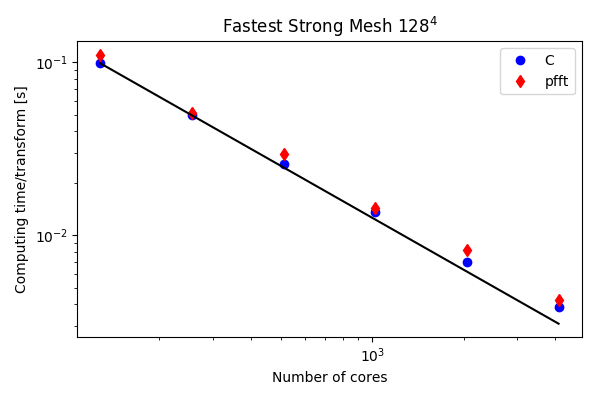}
\caption{Strong scaling of complete real-to-complex FFT on a mesh that
is size $128^4$, using a 3D processor mesh. Showing fastest time
measured.
\label{fig:Strong_4d_complete}}
\end{figure}

Finally, as a proof of concept, we consider a transform of a
4-dimensional array, that can be performed with 3 processor groups as
shown in Sec~\ref{sec:4d}.  We compare in
Fig. \ref{fig:Strong_4d_complete} the strong scaling of one forward
and backward transform to the time used by PFFT on a real array of
size $128^4$. The mesh decomposition is for both codes as chosen by
\MPI{DIMS_CREATE}. Evidently, for this case our C-code is
approximately $5-15\, \%$ faster for the number of processors ranging
from 128 to 4096.\footnote{Note that the global redistribution and serial FFT
times are not shown for this case since we did not manage to get
consistent timings from internal PFFT routines.}

\section{Conclusions}
\label{sec:conclusions}

We presented a new and straightforward approach to implement global
redistributions of arrays required in, but not limited to, parallel
multidimensional fast Fourier transforms. Our approach is based on MPI
subarray datatypes and generalised all-to-all scatter/gather
(\MPI{ALLTOALLW}) collective communications, effectively eliminating
any local array permutations as required in traditional
implementations. Overall, our implementation amounts to less than 100
lines of simple and readable C code taking advantage of high-level
MPI-2 features. Despite such conciseness, our method can perform
global redistributions of $d$-dimensional arrays on up to
$(d-1)$-dimensional process grids between arbitrary pairs of
directions.

We performed a range of strong and weak scaling tests on Shaheen II, a
Cray XC40 system at KAUST, to compare the performance of our method
against other well stablished, mature alternatives like FFTW, PFFT,
P3DFFT, and 2DECOMP\&FFT. Despite \MPI{ALLTOALLW} lacking the
optimizations of other all-to-all collectives, wall-clock time
measurements show that our implementation is on par with the
competitors.

An implementation of the approach discussed in this paper is publicly
available in the open-source Python package
\emph{mpi4py-fft}~\cite{mpi4py-fft}. This package is at the core of
\emph{shenfun}~\cite{shenfun}, a Python-based framework generalizing
the spectral Galerkin method for solving partial differential
equations on tensor-product spaces of arbitrary dimension. Or plans
for future work are geared towards reaching wider audiences by
developing a C library with Fortran wrappers and a software quality
level on par with FFTW.

\section*{Acknowledgments}

M.~Mortensen acknowledges support from the 4DSpace Strategic Research
Initiative at the University of Oslo. L.~Dalcin and D.E.~Keyes
acknowledge support from King Abdullah University of Science and
Technology (KAUST) and the KAUST Supercomputing Laboratory for the use
of the Shaheen supercomputer.

%\section*{References} % arXiv: not needed
\pdfbookmark{References}{references}
\biboptions{numbers,sort&compress}
\bibliographystyle{elsarticle-num}
\bibliography{fft_paper}

\begin{thebibliography}{10}
\expandafter\ifx\csname url\endcsname\relax
  \def\url#1{\texttt{#1}}\fi
\expandafter\ifx\csname urlprefix\endcsname\relax\def\urlprefix{URL }\fi
\expandafter\ifx\csname href\endcsname\relax
  \def\href#1#2{#2} \def\path#1{#1}\fi

\bibitem{foster97}
I.~T. Foster, P.~H. Worley, Parallel algorithms for the spectral transform
  method, SIAM Journal on Scientific Computing 18~(3) (1997) 806--837.
\newblock \href {http://dx.doi.org/10.1137/S1064827594266891}
  {\path{doi:10.1137/S1064827594266891}}.

\bibitem{gupta93}
A.~Gupta, V.~Kumar, The scalability of {FFT} on parallel computers 4~(8) (1993)
  922--932.
\newblock \href {http://dx.doi.org/10.1109/71.238626}
  {\path{doi:10.1109/71.238626}}.

\bibitem{Ding95}
H.~Q. Ding, R.~D. Ferraro, D.~B. Gennery, A portable {3D} {FFT} package for
  distributed-memory parallel architectures, in: {Proceedings of 7th SIAM
  Conference on Parallel Processing}, SIAM Press, 1995, pp. 70--71.

\bibitem{pekurovsky2012}
D.~Pekurovsky, {P3DFFT}: a framework for parallel computations of {F}ourier
  transforms in three dimensions, SIAM Journal on Scientific Computing 34~(4)
  (2012) 192--209.
\newblock \href {http://dx.doi.org/10.1137/11082748X}
  {\path{doi:10.1137/11082748X}}.

\bibitem{Li2010}
N.~Li, S.~Laizet, {2DECOMP\&FFT} - a highly scalable 2d decomposition library
  and {FFT} interface, in: CUG 2010 Proceedings, Cray User Group, 2010.

\bibitem{Pippig13}
M.~Pippig, {PFFT}: An extension of {FFTW} to massively parallel architectures,
  SIAM Journal on Scientific Computing 35~(3) (2013) C213--C236.
\newblock \href {http://dx.doi.org/10.1137/120885887}
  {\path{doi:10.1137/120885887}}.

\bibitem{fftw}
M.~Frigo, S.~G. Johnson, The design and implementation of {FFTW3}, Proceedings
  of the IEEE 93~(2) (2005) 216--231.
\newblock \href {http://dx.doi.org/10.1109/JPROC.2004.840301}
  {\path{doi:10.1109/JPROC.2004.840301}}.

\bibitem{DUY2014}
T.~V.~T. Duy, T.~Ozaki, A decomposition method with minimum communication
  amount for parallelization of multi-dimensional {FFT}s, Computer Physics
  Communications 185~(1) (2014) 153--164.
\newblock \href {http://dx.doi.org/10.1016/j.cpc.2013.08.028}
  {\path{doi:10.1016/j.cpc.2013.08.028}}.

\bibitem{accfft}
A.~Gholami, J.~Hill, D.~Malhotra, G.~Biros, \href{http://accfft.org}{{AccFFT} -
  a new parallel {FFT} library}.
\newline\urlprefix\url{http://accfft.org}

\bibitem{plimptonFFT}
S.~Plimpton,
  \href{{http://www.sandia.gov/\%7Esjplimp/docs/fft/README.html}}{Parallel
  {FFT} package}.
\newline\urlprefix\url{{http://www.sandia.gov/\%7Esjplimp/docs/fft/README.html}}

\bibitem{fftwpp}
J.~C. Bowman, M.~Roberts, \href{http://fftwpp.sourceforge.net}{{FFTW++}: Fast
  {F}ourier transform {C++} header/{MPI} transpose for {FFTW3}}.
\newline\urlprefix\url{http://fftwpp.sourceforge.net}

\bibitem{mpifft4py}
M.~Mortensen, \href{https://github.com/spectralDNS/mpiFFT4py}{Parallel {FFT} in
  {3D} or {2D} using {MPI} for {P}ython}.
\newline\urlprefix\url{https://github.com/spectralDNS/mpiFFT4py}

\bibitem{mpi4py}
L.~Dalcin, \href{https://bitbucket.org/mpi4py/mpi4py/}{{MPI} for {P}ython}.
\newline\urlprefix\url{https://bitbucket.org/mpi4py/mpi4py/}

\bibitem{dalcin08}
L.~Dalcin, R.~Paz, M.~Storti, J.~D'Elia, {MPI} for {P}ython: Performance
  improvements and {MPI}-2 extensions, Journal of Parallel and Distributed
  Computing 68~(5) (2008) 655--662.
\newblock \href {http://dx.doi.org/10.1016/j.jpdc.2007.09.005}
  {\path{doi:10.1016/j.jpdc.2007.09.005}}.

\bibitem{mpistd31}
{Message Passing Interface Forum},
  \href{http://mpi-forum.org/docs/mpi-3.1/mpi31-report.pdf}{MPI: A
  Message-Passing Interface Standard, Version 3.1}, High Performance Computing
  Center Stuttgart (HLRS), 2015.
\newline\urlprefix\url{http://mpi-forum.org/docs/mpi-3.1/mpi31-report.pdf}

\bibitem{hoefler10}
T.~Heofler, S.~Gottlieb, Recent Advances in the Message Passing Interface,
  Springer, 2010, Ch. Parallel Zero-Copy Algorithms for Fast {F}ourier
  Transform and Conjugate Gradient using {MPI} Datatypes, pp. 132--141.
\newblock \href {http://dx.doi.org/10.1007/978-3-642-15646-5}
  {\path{doi:10.1007/978-3-642-15646-5}}.

\bibitem{warp}
D.~Grote, J.-L. Vay, A.~Friedman, S.~Lund,
  \href{https://sites.google.com/a/lbl.gov/warp/home}{Warp}.
\newline\urlprefix\url{https://sites.google.com/a/lbl.gov/warp/home}

\bibitem{fftpack}
P.~N. Swarztrauber, \href{http://www.netlib.org/fftpack/}{{FFTPACK}}.
\newline\urlprefix\url{http://www.netlib.org/fftpack/}

\bibitem{ibm-essl}
{IBM}, \href{http://www.ibm.com/systems/power/software/essl/}{Engineering and
  scientific subroutine library (essl) and parallel essl}.
\newline\urlprefix\url{http://www.ibm.com/systems/power/software/essl/}

\bibitem{intel-mkl}
{Intel}, \href{https://software.intel.com/mkl}{{Math Kernel Library (MKL)}}.
\newline\urlprefix\url{https://software.intel.com/mkl}

\bibitem{petsc-user-ref}
S.~Balay, S.~Abhyankar, M.~F. Adams, J.~Brown, P.~Brune, K.~Buschelman,
  L.~Dalcin, V.~Eijkhout, W.~D. Gropp, D.~Kaushik, M.~G. Knepley, D.~A. May,
  L.~C. McInnes, R.~T. Mills, T.~Munson, K.~Rupp, P.~Sanan, B.~F. Smith,
  S.~Zampini, H.~Zhang, H.~Zhang,
  \href{http://www.mcs.anl.gov/petsc/petsc-current/docs/manual.pdf}{{PETS}c
  users manual}, Tech. Rep. ANL-95/11 - Revision 3.9, Argonne National
  Laboratory (2018).
\newline\urlprefix\url{http://www.mcs.anl.gov/petsc/petsc-current/docs/manual.pdf}

\bibitem{mortensen16}
M.~Mortensen, H.~P. Langtangen, High performance {P}ython for direct numerical
  simulations of turbulent flows, Computer Physics Communications 203 (2016)
  53--65.
\newblock \href {http://dx.doi.org/10.1016/j.cpc.2016.02.005}
  {\path{doi:10.1016/j.cpc.2016.02.005}}.

\bibitem{advmpi}
W.~Gropp, T.~Hoefler, R.~Thakur, E.~Lusk, Using Advanced MPI: Modern Features
  of the Message-Passing Interface, The MIT Press, 2014.

\bibitem{top500}
\href{https://www.top500.org/system/178515}{{Shaheen II CrayXC40 - Top 500 list
  - November 2017}}.
\newline\urlprefix\url{https://www.top500.org/system/178515}

\bibitem{kumar08}
R.~Kumar, A.~Mamidala, D.~K. Panda, Scaling alltoall collective on multi-core
  systems, in: 2008 IEEE International Symposium on Parallel and Distributed
  Processing, IEEE, 2008, pp. 1--8.
\newblock \href {http://dx.doi.org/10.1109/IPDPS.2008.4536141}
  {\path{doi:10.1109/IPDPS.2008.4536141}}.

\bibitem{mpi4py-fft}
L.~Dalcin, M.~Mortensen,
  \href{https://bitbucket.org/mpi4py/mpi4py-fft}{{mpi4py-fft}}.
\newline\urlprefix\url{https://bitbucket.org/mpi4py/mpi4py-fft}

\bibitem{shenfun}
M.~Mortensen, {Shenfun} -- automating the spectral {Galerkin} method, ArXiv
  e-prints\href {http://arxiv.org/abs/1708.03188} {\path{arXiv:1708.03188}}.

\end{thebibliography}

\appendix
\pdfbookmark{Appendix}{appendix}

\clearpage % XXX remove

\section{Full 3D complex FFT with 2D pencil decomposition}
\label{app:3D}
\begin{lstlisting}[
numbers=left,
firstline=10,lastline=89,
basicstyle=\footnotesize\ttfamily,
]
#include <stdlib.h>
#include <assert.h>
#include <complex.h>
#include <math.h>
#include <mpi.h>

// External routine in charge of computing
// multidimensional complex FFTs in-place
enum { FORWARD=-1, BACKWARD=+1 };
extern void seqxfftn(int ndims, int sizes[ndims],
                     double complex *array,
                     int axis, int sign);

// Helper function to compute local sizes
static int lsz(int N, MPI_Comm comm)
{
  int size, rank, n, s;
  MPI_Comm_size(comm, &size);
  MPI_Comm_rank(comm, &rank);
  decompose(N, size, rank, &n, &s);
  return n;
}

// Helper macros
#define product(n)     (n[0]*n[1]*n[2])
#define allocate(t,n)  malloc(product(n)*sizeof(t))
#define deallocate(a)  free(a)

int main(int argc, char *argv[])
{
  MPI_Init(&argc, &argv);

  // Define global 3D array sizes
  int N[3] = {42, 127, 256};

  // Create subgroups from 2D process grid
  MPI_Comm P[2];
  subcomm(MPI_COMM_WORLD, 2, P);

  // Define elementary MPI datatype
  MPI_Datatype T = MPI_C_DOUBLE_COMPLEX;

  // Define local sizes and allocate local 3D arrays
  int sizesA[3] = {lsz(N[0],P[0]), lsz(N[1],P[1]), N[2]};
  int sizesB[3] = {lsz(N[0],P[0]), N[1], lsz(N[2],P[1])};
  int sizesC[3] = {N[0], lsz(N[1],P[0]), lsz(N[2],P[1])};
  double complex *arrayA = allocate(double complex, sizesA);
  double complex *arrayB = allocate(double complex, sizesB);
  double complex *arrayC = allocate(double complex, sizesC);

  for (int j=0, n=product(sizesA); j<n; j++)
    arrayA[j] = j + j * I; // Fill array with complex values

  // Forward FFT
  seqxfftn(3, sizesA, arrayA, 2, FORWARD);
  exchange(P[1], T, 3, sizesA, arrayA, 2, sizesB, arrayB, 1);
  seqxfftn(3, sizesB, arrayB, 1, FORWARD);
  exchange(P[0], T, 3, sizesB, arrayB, 1, sizesC, arrayC, 0);
  seqxfftn(3, sizesC, arrayC, 0, FORWARD);

  // Backward FFT
  seqxfftn(3, sizesC, arrayC, 0, BACKWARD);
  exchange(P[0], T, 3, sizesC, arrayC, 0, sizesB, arrayB, 1);
  seqxfftn(3, sizesB, arrayB, 1, BACKWARD);
  exchange(P[1], T, 3, sizesB, arrayB, 1, sizesA, arrayA, 2);
  seqxfftn(3, sizesA, arrayA, 2, BACKWARD);

  for (int j=0, n=product(sizesA); j<n; j++) // Check result
    assert(fabs(creal(arrayA[j]) - j) < 1.0e-8 &&
           fabs(cimag(arrayA[j]) - j) < 1.0e-8);

  deallocate(arrayA);
  deallocate(arrayB);
  deallocate(arrayC);
  MPI_Comm_free(&P[0]);
  MPI_Comm_free(&P[1]);

  MPI_Finalize();
  return 0;
}
\end{lstlisting}

\section{Full 4D complex FFT with 3D decomposition}
\label{app:4D}
\begin{lstlisting}[
numbers=none,
firstline=10,lastline=103,
basicstyle=\footnotesize\ttfamily,
]
#include <stdlib.h>
#include <assert.h>
#include <complex.h>
#include <math.h>
#include <mpi.h>

// External routine in charge of computing
// multidimensional complex FFTs in-place
enum { FORWARD=-1, BACKWARD=+1 };
extern void seqxfftn(int ndims, int sizes[ndims],
                     double complex *array,
                     int axis, int sign);

// Helper function to compute local sizes
static int lsz(int N, MPI_Comm comm)
{
  int size, rank, n, s;
  MPI_Comm_size(comm, &size);
  MPI_Comm_rank(comm, &rank);
  decompose(N, size, rank, &n, &s);
  return n;
}

// Helper macros
#define product(n)     (n[0]*n[1]*n[2]*n[3])
#define allocate(t,n)  malloc(product(n)*sizeof(t))
#define deallocate(a)  free(a)

int main(int argc, char *argv[])
{
  MPI_Init(&argc, &argv);

  // Define global 4D array sizes
  int N[4] = {16, 17, 18, 19};

  // Create subgroups from 3D process grid
  MPI_Comm P[3];
  subcomm(MPI_COMM_WORLD, 3, P);

  // Define elementary MPI datatype
  MPI_Datatype T = MPI_C_DOUBLE_COMPLEX;

  // Define local sizes
  int sizesA[4] = {lsz(N[0],P[0]), lsz(N[1],P[1]),
                   lsz(N[2],P[2]), N[3]};
  int sizesB[4] = {lsz(N[0],P[0]), lsz(N[1],P[1]),
                   N[2], lsz(N[3],P[2])};
  int sizesC[4] = {lsz(N[0],P[0]), N[1],
                   lsz(N[2],P[1]), lsz(N[3],P[2])};
  int sizesD[4] = {N[0], lsz(N[1],P[0]),
                   lsz(N[2],P[1]), lsz(N[3],P[2])};

  // Allocate local 3D arrays
  double complex *arrayA = allocate(double complex, sizesA);
  double complex *arrayB = allocate(double complex, sizesB);
  double complex *arrayC = allocate(double complex, sizesC);
  double complex *arrayD = allocate(double complex, sizesD);

  for (int j=0, n=product(sizesA); j<n; j++)
    arrayA[j] = j + j * I; // Fill array with complex values

  // Forward FFT
  seqxfftn(4, sizesA, arrayA, 3, FORWARD);
  exchange(P[2], T, 3, sizesA, arrayA, 3, sizesB, arrayB, 2);
  seqxfftn(4, sizesB, arrayB, 2, FORWARD);
  exchange(P[1], T, 3, sizesB, arrayB, 2, sizesC, arrayC, 1);
  seqxfftn(4, sizesC, arrayC, 1, FORWARD);
  exchange(P[0], T, 3, sizesC, arrayC, 1, sizesD, arrayD, 0);
  seqxfftn(4, sizesD, arrayD, 0, FORWARD);

  // Backward FFT
  seqxfftn(4, sizesD, arrayD, 0, BACKWARD);
  exchange(P[0], T, 3, sizesD, arrayD, 0, sizesC, arrayC, 1);
  seqxfftn(4, sizesC, arrayC, 1, BACKWARD);
  exchange(P[1], T, 3, sizesC, arrayC, 1, sizesB, arrayB, 2);
  seqxfftn(4, sizesB, arrayB, 2, BACKWARD);
  exchange(P[2], T, 3, sizesB, arrayB, 2, sizesA, arrayA, 3);
  seqxfftn(4, sizesA, arrayA, 3, BACKWARD);

  for (int j=0, n=product(sizesA); j<n; j++) // Check result
    assert(fabs(creal(arrayA[j]) - j) < 1.0e-8 &&
           fabs(cimag(arrayA[j]) - j) < 1.0e-8);

  deallocate(arrayA);
  deallocate(arrayB);
  deallocate(arrayC);
  deallocate(arrayD);
  MPI_Comm_free(&P[0]);
  MPI_Comm_free(&P[1]);
  MPI_Comm_free(&P[2]);

  MPI_Finalize();
  return 0;
}
\end{lstlisting}

\end{document}